\documentclass[prd,showpacs,twocolumn,nofootinbib,superscriptaddress]{revtex4}
\usepackage{amsmath}
\usepackage{multirow}
\usepackage{pstricks}

\newcommand{\dsl}{\partial\hspace{-1.8mm}/}
\newcommand{\psl}{p\hspace{-1.7mm}/}

\newcommand{\ave}[1]{\langle {#1} \rangle}
\newcommand{\pb}{\bar q}
\newcommand{\qq}{\ave{\pb q}}

\newcommand{\Tr}{{\mathrm{\rm Tr}}}
\newcommand{\intdk}{\int\frac{d^4k}{(2 \pi)^4}}
\newcommand{\intdEk}{\int\frac{d_E^4k}{(2 \pi)^4}}
\newcommand{\intdl}{\int\frac{d^4l}{(2 \pi)^4}}

\usepackage{graphicx,epsfig}

\begin{document}
\title{Nonlocal PNJL model beyond mean field and the QCD phase transition}

\author{A.E.~Radzhabov}
\email{aradzh@icc.ru}
\affiliation{Institute for System Dynamics and Control Theory, 664033 Irkutsk,
Russia}

\author{D. Blaschke}
\email{blaschke@ift.uni.wroc.pl}
\affiliation{Institute for Theoretical Physics, University of Wroc{\l}aw, 50-204 Wroc{\l}aw, Poland}
\affiliation{Bogoliubov  Laboratory of Theoretical Physics, JINR Dubna, 141980 Dubna, Russia}

\author{M. Buballa}
\email{michael.buballa@physik.tu-darmstadt.de}
\affiliation{Institut f\"ur Kernphysik, Technische Universit\"at Darmstadt,
D-64289 Darmstadt,
Germany}

\author{M.K.~Volkov}
\email{volkov@theor.jinr.ru}
\affiliation{Bogoliubov  Laboratory of Theoretical Physics, JINR Dubna, 141980 Dubna, Russia}

\begin{abstract}

A nonlocal chiral quark model is consistently extended beyond mean field using
a strict $1/N_c$ expansion scheme.
The parameters of the nonlocal model are refitted so that the physical values of the pion mass and the weak pion decay constant are obtained.
The size of the $1/N_c$ correction to the quark condensate is carefully studied
and compared with the usual local Nambu--Jona-Lasinio model.
It is found that even the sign of the corrections can be different.
This can be attributed to the mesonic cut-off of the local model.
The model is also applied to finite temperature. We find that the $1/N_c$
corrections dominate the melting of the chiral condensate at low
temperatures, $T\lesssim$~100~MeV, in agreement with chiral perturbation
theory. On the other hand, the relative importance of the $1/N_c$ corrections
in the cross-over regime depends on the parameter  $T_0$  of the Polyakov
loop potential. For $T_0=270$~MeV, corresponding to a fit of lattice data for
pure gluodynamics, the correction terms are large and lead to a lowering
of the chiral phase transition temperature in comparison with the mean-field
result. Near the phase transition the $1/N_c$ expansion breaks down and a
non-perturbative scheme is needed to include mesonic correlations in that
regime. Lowering $T_0$ leads to a more rapid cross-over even at the mean-field
level and the unstable region for the $1/N_c$ corrections shrinks.
For $T_0 \lesssim 220$ MeV the temperatures of deconfinement and chiral
restoration are practically synchronized.

\end{abstract}
\maketitle

\section{Introduction}

A quantum field theoretical description of strong interactions in the
nonperturbative regime is one of the most interesting and challenging problems
of present-day theoretical physics.
Quantum chromodynamics is well known only at the perturbative level
whereas the low-energy region and the most interesting ``hadronic'' phase
in the QCD phase diagram is in the nonperturbative regime.
The only nonperturbative \textit{ab initio} calculations are performed in
lattice QCD, but their range of applicability is still limited.
To gain some analytical insights to nonperturbative QCD, continuum approaches,
even using effective models, are legitimate tools.
They may provide a theoretical interpretation of results from lattice QCD and
allow their extrapolation to otherwise inaccessible domains.

One of the successful models for a description of chiral quark dynamics and
the phase diagram is the Nambu--Jona-Lasinio (NJL) model \cite{Nambu:1961tp}
applied to quarks
\cite{Volkov:1984kq,Klimt:1989pm,Klevansky:1992qe,Hatsuda:1994pi}.
This model provides a mechanism for spontaneous chiral symmetry breaking and
the formation of a quark condensate.
The low-lying hadron spectrum, low-energy dynamics, the main strong and
electromagnetic decays, hadron-hadron scattering and the internal
characteristics of mesons have a reasonable explanation within this model.
A generalization of the NJL model has been proposed which includes
the coupling of the chiral quark sector to the Polyakov loop, being an order
parameter of the deconfinement transition
(PNJL model \cite{Meisinger:1995ih,Fukushima:2003fw,Megias:2004hj,Ratti:2005jh,
Roessner:2006xn,Sasaki:2006ww,Hansen:2006ee,Blaschke:2007np,Contrera:2007wu,
Hell:2009by}).
The temperature dependent parameters of the effective Polyakov loop potential
are fitted by using lattice QCD data on pure gluodynamics.

Nonlocal generalizations of the PNJL model provide an approach to the
4-momentum dependence of the quark mass function and wave function
renormalization of the quark propagator \cite{Blaschke:2007np,Contrera:2007wu,Hell:2009by,Horvatic:2010md,Kondo:2010ts}, thus allowing to implement
detailed nonperturbative information about low-energy QCD dynamics accessible,
e.g., in ab-initio LQCD simulations  \cite{Kamleh:2007ud,Bazavov:2010bx,Borsanyi:2010bp}.
Recently, a justification of nonlocal PNJL models as effective low-energy
limit of QCD has been given \cite{Kondo:2010ts}, based in part on methods of
the Wilsonian renormalization group \cite{Braun:2009gm}.

Usually, NJL and PNJL models are formulated at the mean-field level.
However, there are physical problems where the mean-field formulation is not
sufficient.
Large corrections to the mean-field behavior can be expected, e.g., in the
description of broad resonances from their coupling to intermediate meson
states\footnote{The $\sigma$-meson with a decay width of the order of its mass
is the most striking example of such state. See, e.g., ``Note on scalar
mesons'' in \cite{Amsler:2008zzb}.}
and for the equation of state of the hadronic phase where quark and
gluon degrees for freedom are ``frozen'' in condensates and hadronic bound
states are carrying all dynamics and thus represent the physical degrees of
freedom.

There are different schemes to go beyond the mean-field level
\cite{Quack:1993ie,Ebert:1994sm,Nikolov:1996jj,Dmitrasinovic:1995cb,Blaschke:1995gr,Oertel:1999fk,Oertel:2000jp,Plant:2000ty,Jafarov:2003pe,Goeke:2007bj,Muller:2010am}.
One of the most promising ones is based on a strict
expansion of the inverse number of quark colors, $1/N_c$, which is a natural
expansion parameter for gauge theories \cite{'tHooft:1973jz}.
The local NJL model is nonrenormalizable and therefore it is necessary to
introduce an additional cut-off parameter when going beyond the mean-field
level.
This problem is absent in nonlocal versions of the NJL model where the
nonlocality leads to an effective regularization which renders the quark
(multi)-loop diagrams convergent.

In the present paper the $SU(2) \times SU(2)$ nonlocal PNJL model is
investigated beyond mean field  within a strict $1/N_c$ expansion scheme
at finite temperature and zero chemical potential.
This study aims to quantify the effect of mesonic excitations on
the chiral restoration temperature and to demonstrate that the behavior of
the chiral condensate at low-temperatures is in accordance with the exact
results of chiral perturbation theory.
We also investigate the dependence of the results on the parameter $T_0$ of the
Polyakov-loop potential.

\section{Nonlocal model in vacuum}

We begin with the discussion of our nonlocal model in vacuum.
The main goal is to fix the model parameters by calculating meson
properties at next-to-leading order in $1/N_c$.
It is well-known that in the local PNJL model at $T=0$ the gluon sector
decouples from the quark sector, so that the latter is reduced to the
standard NJL model. The same is true in the nonlocal model.
In this section we can therefore restrict ourselves to the quark
sector, while the Polyakov-loop dynamic will be introduced in
Sec.~\ref{sec:T}.

\subsection{Mean field}

The quark sector of the nonlocal chiral quark model is described by the
Lagrangian
\begin{eqnarray}
\mathcal{L}_q&=& \bar{q}(x)(i \dsl -m_c)q(x)
+\frac{G}{2}[J_\sigma^2(x) + \vec J^{\,2}_\pi(x)] ~,\label{QLagrangian}
\end{eqnarray}
where $m_c$ is the current quark mass.
The nonlocal quark currents are
\begin{eqnarray}
J_\mathrm{M}(x)&=&\int d^4x_1 d^4x_2 ~f(x_1)f(x_2)\times\nonumber\\
&&\quad\times\bar{q}(x-x_1)\mathbf{\Gamma}_\mathrm{M} q(x+x_2),
\label{QCurrents}
\end{eqnarray}
where $\mathbf{\Gamma}_\mathrm{\sigma}= \mathrm{1}$,
$\mathbf{\Gamma}_\mathrm{\pi}= i \gamma^5 \tau^a$ with $a=1,2,3$,
and $f(x)$ is a form factor.
The latter is defined by its Fourier transform in Euclidean
space\footnote{Unless stated otherwise, the expressions in this
paper are given in Minkowski space.
The transformation to Euclidean space is trivial for energies below the
(pseudo)threshold (see Sect. \ref{VacuumResults}).}, which we take
to be Gaussian, $f^2(p_E^2)=\exp(-p_E^2/\Lambda^2)$.
This scheme introducing nonlocal currents which we utilize throughout this
work has been denoted as ``instanton liquid model'' as opposed to the
``one-gluon exchange model'' scheme, see Ref.~\cite{GomezDumm:2006vz} for
details.

After linearization of the four-fermion vertices by introducing auxiliary
scalar  ($\tilde{\sigma}$) and pseudoscalar ($\pi^a$) meson fields the quark
sector is described by the Lagrangian\footnote{Note, that after bosonization
the possible exchange (Fock) terms between quark currents are eliminated.}
\begin{eqnarray}
   \mathcal{L}_{q\pi\sigma} &=& \bar{q}(x)(i \dsl -m_c)q(x)
   - \frac{\pi_a^2(x)+\tilde{\sigma}^2(x)}{2G}+\nonumber\\
    &&+ J_\sigma(x)\tilde{\sigma}(x)+\pi^a(x)J_\pi^a(x)~.
\label{lagr2}
\end{eqnarray}
To proceed, we single out the nonzero mean-field value of the scalar field by
the decomposition $\tilde{\sigma}=\sigma+\sigma_{\rm MF}$ so that $\pi^a$ and
$\sigma$ denote only the fluctuating parts of the fields
($\langle\pi^a \rangle=\langle \sigma\rangle=0$)
describing mesonic correlations. The scalar mean field gives a dynamical
contribution to the quark mass, i.e., the dressed quark propagator
becomes
\begin{eqnarray}
S(p)&=&\left(\psl - m(p^2)\right)^{-1}
     = \left(\psl - m_c - \Sigma(p)\right)^{-1},
\label{Smf}
\end{eqnarray}
with the quark self-energy
\begin{eqnarray}
\Sigma(p)
&=& i G \Gamma^\sigma(p,p) \intdk \;\Tr\,[\,\Gamma^\sigma(k,k)\,
S(k)\,]\nonumber\\
&\equiv& m_d f^2(p^2)~.
\label{Sigmamf}
\end{eqnarray}
Here the symbol $\Tr$ stands for the trace over color-, flavor- and
Dirac-indices, and
$\Gamma^\mathrm{M}(q_1,q_2) = \mathbf{\Gamma}_\mathrm{M} f(q_1^2)f(q_2^2)$
is the nonlocal generalization of the meson-quark-antiquark vertex function
with the quark and antiquark momenta  $q_1$ and $q_2$, respectively.
In the following, we will often use a shorthand subscript notation for
the momentum dependence of functions, e.g., $f_{k} \equiv f(k^2)$ and
$\Gamma^\sigma(q_1,q_2) \equiv \Gamma^\sigma_{q_1,q_2}$.

The amplitude $m_d=-\sigma_{\rm MF}$ is an order parameter for dynamical chiral
symmetry breaking. The chiral condensate per flavor,
\begin{eqnarray}
     \qq^{\mathrm{MF}}  &=&
     - \frac{i}{2} \intdk \;\Tr\,[\, S^{np}(k) \,],
\label{qbarq}
\end{eqnarray}
is obtained from the non-perturbative part of the quark propagator,
$S^{np}(p)=S(p) - S^c(p)$, i.e., after subtracting the perturbative part
$S^c(p)=(p\hspace{-1.8mm}/ - m_c)^{-1}$.

Mesons are described as bound state solutions of the quark-antiquark
Bethe-\-Salpeter equation.
The meson propagators are given by
\begin{eqnarray}
\mathrm{D}^{\mathrm{M}}_p=\frac{1}{-G^{-1}+\Pi^{\mathrm{M}}_p}
\label{mesonprop},
\end{eqnarray}
where $\mathrm{M}=\pi,\sigma$ and
$\Pi^{\mathrm{M}}_p \equiv \Pi_{\mathrm{M}}^{\mathrm{MF}}(p^2) $
are the mean field polarization functions defined by
\begin{eqnarray}
\Pi^{\mathrm{M}}_p= i  \int \frac{d^4k}{(2\pi)^4}
\Tr \left[ S_{k_-}\Gamma^{\mathrm{M}}_{k_-,k_+} S_{k_+}
\Gamma^{\mathrm{M}}_{k_+,k_-} \right],
\label{poloper}
\end{eqnarray}
where $k_\pm=k \pm p/2$.

The meson masses are the poles of the propagators at
$p^2=(M_\mathrm{M}^\mathrm{MF})^2$ obtained by solving
\begin{eqnarray}
-G^{-1}+\Pi_{\mathrm{M}}^{\mathrm{MF}}\left((M_\mathrm{M}^\mathrm{MF})^2\right)=0.
\end{eqnarray}
Let us consider the pion case.
In the vicinity of the pole the pion propagator can be expanded as
\begin{eqnarray}
\mathrm{D}^{\pi}_p
   \simeq \frac{\left(g^\mathrm{MF}_\pi\right)^2}{p^2-(M_\pi^\mathrm{MF})^2}
+ \mathrm{regular\,terms}~,
\label{pionpole}
\end{eqnarray}
where $g^\mathrm{MF}_\pi$ is the pion-quark-antiquark coupling constant
\begin{eqnarray}
\left.\left(g^\mathrm{MF}_\pi\right)^{-2}
=\frac{\partial \Pi_\pi^\mathrm{MF}(p^2)}
{\partial p^2 }\right|_{p^2=(M_\pi^\mathrm{MF})^2}
\label{mesonmasses}
\end{eqnarray}

\begin{figure} 
\begin{tabular}{ccccc}
\includegraphics[height=0.07\textheight]{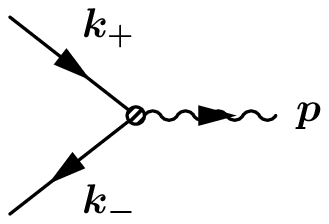} &\hspace{0.08\textwidth}&
\includegraphics[height=0.07\textheight]{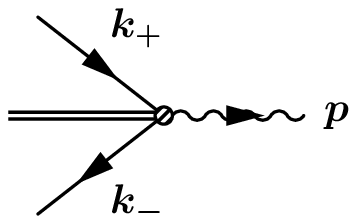}
\\ $\Gamma^{5}_{k_-,k_+}$ && $\Gamma^{5M}_{k_-,k_+,p}$
\end{tabular}
  \caption{Vertices of weak currents (wavy lines) coupled to a quark (left)
           and to the non-local quark-meson vertices (right).}
    \label{WeakCurrentVertices}
\end{figure}

To calculate the weak pion decay constant, the pionic solutions of the
Bethe-\-Salpeter equation have to be coupled to an external weak current.
To that end the Lagrangian (\ref{lagr2}) must be modified so that it
becomes invariant under local vector and axial-vector gauge
transformations.
In nonlocal models this is complicated by the fact that not only
the kinetic part part but also the interaction is not
gauge invariant by itself, so that in addition to the usual covariant
derivative one has to take into account the coupling of the external
fields to the nonlocal quark vertices.
This can be done by delocalization of the quark fields~\cite{Terning:1991yt,Bowler:1994ir,Plant:1997jr,Dorokhov:2003kf,Scarpettini:2003fj,GomezDumm:2006vz}
\begin{eqnarray}
q(y)\to Q(x,y) = E(x,y)q(y)
\end{eqnarray}
where
\begin{eqnarray}
E(x,y)= \mathcal{P}\mathrm{exp}\left\{i\int
\limits_x^y dz^\mu [{\cal V}^a_\mu(z)+ {\cal A}^a_\mu(z)\gamma_5]T^a\right\}
\label{Schwinger}
\end{eqnarray}
is the Schwinger phase factor, involving the external vector and
axial-vector gauge fields ${\cal V}^a_\mu$ and ${\cal A}^a_\mu$,
and $T^a\equiv \tau^a/2$.
For the kinetic part it can be shown that this replacement
is equivalent to minimal substitution, hence leading to the standard
electroweak vertices.
In the interaction Lagrangian, on the other hand,
the nonlocal quark currents Eq.~(\ref{QCurrents}) are replaced by
\begin{eqnarray}
J_\mathrm{M}(x)&=&\int d^4x_1 d^4x_2 ~f(x_1)f(x_2)\times\nonumber\\
&&\quad\times\bar{Q}(x-x_1,x)\mathbf{\Gamma}_\mathrm{M} Q(x,x+x_2),
\end{eqnarray}
giving rise to additional vertex contributions.
One of them is related to the coupling of $J_\mathrm{\sigma}$ to
the scalar mean-field $\sigma_{\rm MF}$ and contributes, together
with the bare vertex extracted from the kinetic part, to the
vertex function depicted on the left of Fig.~\ref{WeakCurrentVertices}.
The other contributions originate from  $J_\mathrm{M}$ coupled to the
fluctuating meson fields and lead to the vertex functions shown on the
right of Fig.~\ref{WeakCurrentVertices}.

Although the above procedure is sufficient to ensure gauge
invariance, in general it does not unambiguously fix the vertex structure.
To this end, it is necessary to define rules for the evaluation of
the line integral in Eq.~(\ref{Schwinger}).
This can be done by specifying the integration path, e.g., as a straight
line~\cite{Bowler:1994ir,Plant:1997jr,Scarpettini:2003fj,GomezDumm:2006vz},
or by making use of the path independent definition of the
derivative of the line integral~\cite{Terning:1991yt,Dorokhov:2003kf}.
However, for the pion decay constant we only need the longitudinal projection
of the vertices, which are related to axial Ward-Takahashi identities
and therefore do not depend on the integration path.
The results read
\begin{eqnarray}
\Gamma^{5,L}_{k_-,k_+}   &=&-\frac{1}{p^2}\biggl(S_{k_-}^{-1}\gamma_5 T^a +\gamma_5 T^a S_{k_+}^{-1}+\biggr.\nonumber\\
&&\biggl.+2(m_c+ m_d f_{k_-} f_{k_+}) \gamma_5 T^a \biggr)~,\nonumber\\
\Gamma^{5\mathrm{M},L}_{k_-,k_+}&=&-\frac{1}{p^2}\biggl[\left(f_{k_+}- f_{k_+-p} \right)f_{k_-}\gamma_5 T^a \mathbf{\Gamma}_\mathrm{M} +\biggr.\nonumber\\
&&\biggl.+ \left(f_{k_-}- f_{k_-+p} \right)f_{k_+} \mathbf{\Gamma}_\mathrm{M}\gamma_5 T^a\biggr].
\end{eqnarray}

\begin{figure} 
\begin{tabular}{ccccc}
\includegraphics[width=0.15\textwidth]{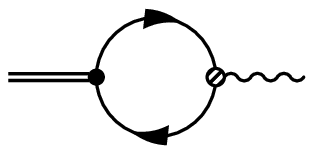} &\hspace{0.08\textwidth}&
\includegraphics[width=0.15\textwidth]{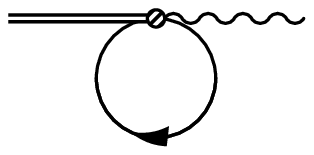}
\\ 1 && 2
\end{tabular}
  \caption{Weak pion decay at mean field.}
    \label{fpMf}
\end{figure}

Accordingly, the pion decay constant at the mean-field level contains two
pieces,
$f_{\pi}^{\rm MF}(p^2)=f_{\pi,1}^{\rm MF}(p^2)+f_{\pi,2}^{\rm MF}(p^2)$,
which are displayed in Fig.~\ref{fpMf}.
Evaluating these diagrams one finds
\begin{eqnarray}
f_{\pi,1}^{\mathrm{MF}}  (p^2)&=&
  g_{\pi}^{\mathrm{MF}} \intdk
  \Tr \left[  \Gamma^{\pi}_{k_+,k_-}  S_{k_-} \Gamma^{5,L}_{k_-,k_+} S_{k_+}\right]\nonumber\\
f_{\pi,2}^{\mathrm{MF}}  (p^2)&=&
  g_{\pi}^{\mathrm{MF}} \intdk
   \Tr \left[  \Gamma^{5\pi ,L}_{k,k} S_k \right],
\label{FpiMF}
\end{eqnarray}

With the definition of the pion mass and weak decay constant two of the three
parameters ($m_c$, $\Lambda$, $G\Lambda^2$) of the nonlocal model can be fixed
with the vacuum values of these observables.

\subsection{$1/N_c$ corrections beyond mean field}

As usual in the systematic $1/N_c$ expansion, the four-quark coupling constant
$G$ is considered to be of the order $1/N_c$.
In this case the $N_c$ behavior of pion properties calculated in the model
coincides with leading-order QCD calculations ($f_\pi \sim \sqrt{N_c}$).
As a result any meson propagator line in diagrams has a $1/N_c$ suppression
factor.

\begin{figure} 
\begin{tabular}{ccccc}
        \includegraphics[width=0.1\textwidth]{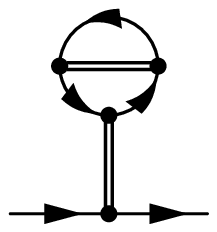}&\hspace{0.1\textwidth}&
        \includegraphics[width=0.15\textwidth]{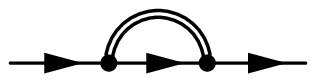}\\
        a && b
\end{tabular}
  \caption{$1/N_c$ corrections to the quark selfenergy.}
    \label{Qucorrection}
\end{figure}

The next-to-leading $1/N_c$ correction $\Sigma^{\mathrm{N_c}}$
to the quark selfenergy corresponds to the diagrams displayed in
Fig.~\ref{Qucorrection}~\cite{Blaschke:1995gr,Oertel:1999fk,Oertel:2000jp}.
From these diagrams one obtains
\begin{eqnarray}
\Sigma^{\mathrm{N_c}}_p =
 \mathrm{C}\, f^2_p
 -  \sum\limits_{\mathrm{M}=\sigma,\pi}
     i\intdl
     \left[\mathrm{D}^\mathrm{M}_l \Gamma^\mathrm{M}_{p,p-l} S_{p-l}
     \Gamma^\mathrm{M}_{p-l,p}\right],\nonumber
\end{eqnarray}
where
\begin{eqnarray}
\mathrm{C} =
    \mathrm{D}^\sigma_0 \zeta,\quad\zeta=\sum\limits_{\mathrm{M}
=\sigma,\pi}i\intdl \mathrm{D}^\mathrm{M}_l \Gamma^{\sigma \mathrm{MM}}_{l,l}~,
\end{eqnarray}
and $\Gamma^{\alpha\beta\gamma}_{q_1,q_2}$ is the quark triangle diagram for
the three-meson vertex.
For external mesons $\alpha$, $\beta$ and $\gamma$ the quark triangle has the
following form
\begin{eqnarray}
 \Gamma^{\alpha\beta\gamma}_{q_1,q_2} &=&
 - i \intdk \Tr \times\,\\
 &\times&\left[\Gamma^{\alpha}_{k+q_1,k+q_2} S_{k+q_2}\Gamma^{\beta}_{k+q_2,k}
S_k \Gamma^{\gamma}_{k,k+q_1} S_{k+q_1}\right].
\nonumber
\label{trianglevertex}
\end{eqnarray}
When $q_1=q_2=l$, the three-meson vertices in Euclidean space are
\begin{eqnarray}
 \Gamma^{\sigma \mathrm{MM}}_{l,l} & =&
  -4 N_c N_f \intdEk \frac{f^2_k f^4_{k+l}}{D_k^{} D^2_{k+l}}\times
\label{trianglevertexll}\\
  &\times&\left[2k\cdot(k+l)m_{k+l}^{}\pm m_k^{} \left((k+l)^2+m^2_{k+l}\right)
\right]~,
\nonumber
\end{eqnarray}
where the upper sign corresponds to $M=\sigma$, the lower sign to $M=\pi$
and $D_k=k^2+m_k^2$ is the denominator of the quark propagator.

From the $1/N_c$ corrections to the quark selfenergy it follows
for the quark propagator
\begin{eqnarray}
 S^{\mathrm{MF+N_c}}_p&=&
\left(S_p^{-1}-\Sigma^{\mathrm{N_c}}_p\right)^{-1}
\nonumber\\
&=& S_p + S_p\Sigma^{\mathrm{N_c}}_p S_p + \text{higher orders},
\end{eqnarray}
which, in turn, gives rise to $1/N_c$ corrections to the quark
condensate (cf.~Eq.~(\ref{qbarq})).
According to the two selfenergy corrections shown in
Fig.~\ref{Qucorrection}, there are two contributions,
$\qq^{\mathrm{N_c}}=\qq^{\mathrm{N_c},a}+\qq^{\mathrm{N_c},b}$,
which are given by
\begin{eqnarray}
\qq^{\mathrm{N_c},a} &=& -\mathrm{C} \Pi^{\mathrm{\sigma},a}_0~, \quad
\qq^{\mathrm{N_c},b} = \zeta^{b}~,
\end{eqnarray}
where $\Pi^{\mathrm{\sigma},a}_p$ can be obtained from Eq.~(\ref{poloper})
for $\Pi^{\mathrm{\sigma}}_p$ by substituting the form factors in the numerator
$f^2_{k_+} f^2_{k_-} \rightarrow f_{k_+}^{} f_{k_-}^{}$ and $\zeta^{b}$ can
be obtained from $\zeta$ by substituting the form factors in the numerator of
the function $\Gamma^{\sigma \mathrm{MM}}_{l,l}$, Eq.~(\ref{trianglevertexll}),
$f^4_{k+l} \rightarrow f^2_{k+l}$ .

The full pion propagator consists of a mean-field part plus $1/N_c$ corrections
(details are given in the appendix)
\begin{eqnarray}
\Pi_{\pi}^{\mathrm{Full}}(p^2)=
\Pi_{\pi}^\mathrm{MF}(p^2)+\Pi_{\pi}^{\mathrm{N_c}}(p^2)
\end{eqnarray}
and the pion mass can be found by solving the equation
\begin{eqnarray}
-G^{-1}+\Pi_{\pi}^{\mathrm{Full}}((M_{\pi}^{\mathrm{Full}})^2)=0.
\end{eqnarray}
The quark-meson coupling constant should be divided into mean-field part and
a $1/N_c$ correction.
This separation is unique only for the pion in the chiral limit.
At finite $m_c$ there are different possibilities to take into account higher
order $p^2$ terms.
In the present work we use
\begin{eqnarray}
(g_\pi^{\mathrm{Full}})^{-2}&=&
\left.\frac{\partial \Pi_{\pi}^{\mathrm{Full}}(p^2)}{\partial p^2 }
\right|_{p^2 = (M_{\pi}^{\mathrm{Full}})^2}~,\nonumber\\
(g_\pi^\mathrm{MF})^{-2}&=&
\left.\frac{\partial \Pi^\mathrm{MF}_{\pi}(p^2)}{\partial p^2 }
\right|_{p^2 = (M^\mathrm{MF}_{\pi})^2}~.
\end{eqnarray}

There are two sources of $1/N_c$ corrections for the weak pion decay constant.
One is the correction of the meson-quark coupling constant and can be obtained
from the mean-field expression, Eq. (\ref{FpiMF}), by the substitution
$g_\pi^\mathrm{MF} \rightarrow g_\pi^{\mathrm{N_c}}=
g_\pi^{\mathrm{Full}}-g_\pi^\mathrm{MF}$.
The other correction is due to new diagrams appearing at order $1/N_c$.
We present details of the calculation of this correction in the appendix.

\subsection{Vacuum results
\label{VacuumResults}}

\begin{figure}[htb]
    \centering
        \includegraphics[width=0.45\textwidth]{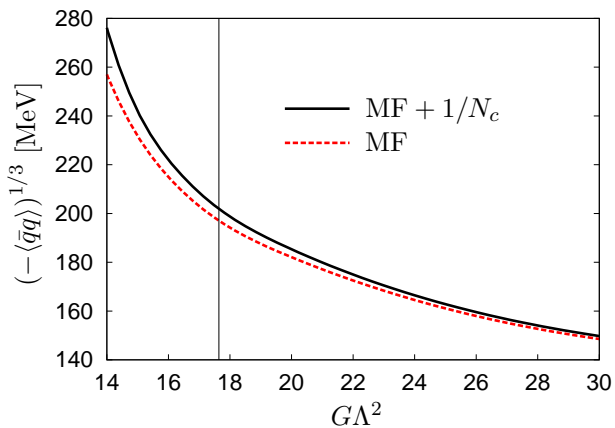}
  \caption{(Color online) The quark condensate as a function of the
dimensionless coupling $G\Lambda^2$. The result of the full $1/N_c$ corrected
approach (black solid line) is compared to the mean field contribution
(red dashed line). The vertical line corresponds to the parameter set No.~4.
  }
    \label{condensate}
\end{figure}

The $1/N_c$ corrections to meson properties will affect the results for the
quark condensate via the readjustment of the model parameters
($\Lambda$, $m_c$, $G\Lambda^2$) which are to be chosen such that the physical
values for the pion mass $M_{\pi^\pm}=139.57$ MeV and the weak pion decay
constant $f_\pi=92.42$ MeV are obtained at $T=0$, while the dimensionless
coupling $G\Lambda^2$ is left as a free parameter.
Different parameterizations of the nonlocal model beyond mean field are given
in Table \ref{FitModParams}.
The corresponding behavior of the quark condensate as a function of the
dimensionless coupling is presented in Fig.~\ref{condensate}.

\begin{figure}[htb]
    \centering
        \includegraphics[width=0.45\textwidth]{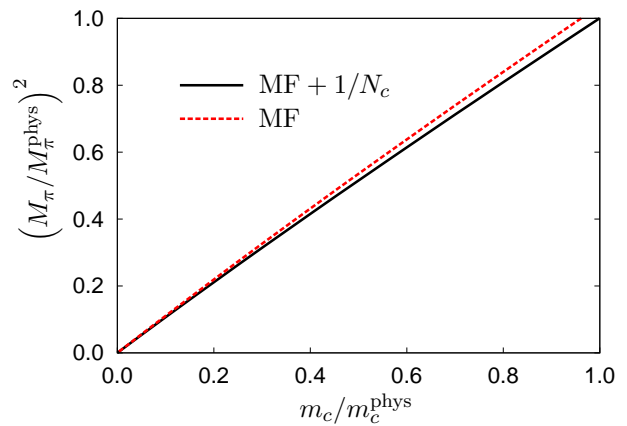}
  \caption{(Color online) Squared pion mass normalized to the physical value
versus normalized current quark mass in mean-field approximation
(red dashed line) and the full $1/N_c$ corrected approach (black solid line).
The values of $\Lambda$ and $G$ correspond to parameter set No.~4.}
  \label{ChiralCheck}
\end{figure}

One important check of the calculations is the proof of the Goldstone
nature of the pion.
In the case of exact chiral symmetry the pion should be massless.
This check is shown in Fig. \ref{ChiralCheck} where $M_{\pi}^{\mathrm{Full}}$
and $M_{\pi}^{\mathrm{MF}}$ are displayed as functions of the current quark
mass for $\Lambda$ and $G$ taken from parameter set 4.

\begin{table*}
\caption{
Different parameterizations: model parameters
($\Lambda$,$G\Lambda^2$,$m_c$), the dynamical mass $m_d$,
position of the first two poles and the corresponding (pseudo-) threshold of
the mean-field quark propagator,
mean-field values of the pion mass and the weak decay constant, and
critical temperatures for chiral restoration and deconfinement for two values of the parameter $T_0$ of Polyakov loop potential.
The set with the highest threshold value is highlighted.
The pseudo-threshold values are shown in italics.}
\begin{tabular}{|ccccccccc|ccc|ccc|}
\hline
\multirow{2}*{No.}&\multirow{2}*{$G\Lambda^2$}&\multirow{2}*{$\Lambda$}&\multirow{2}*{$m_c$}&\multirow{2}*{$m_d$}& \multirow{2}*{first two poles} &\multirow{2}*{(pseudo)threshold}&\multirow {2}*{$M_\pi^{\mathrm{MF}}$}&\multirow {2}*{$f_\pi^{\mathrm{MF}}$}&\multicolumn{3}{|c|}{$T_0=270$ MeV}&\multicolumn{3}{|c|}{$T_0=208$ MeV}\\\cline{10-12}\cline{13-15}
                  &                           &                        &                    &                    &                                &                                &                                     &                                     &$T_c$              &$T_c^{\mathrm{MF}}$&    $T_d$ &$T_c$              &$T_c^{\mathrm{MF}}$&    $T_d$ \\
units     &              &  MeV         &  MeV        &   MeV        & GeV$^2$                      & MeV                &     MeV             &     MeV             &     \multicolumn{3}{|c|}{MeV} &  \multicolumn{3}{|c|}{MeV} \\
\hline
        1 &        13.35 &       1479.2 &        2.82 &        139.2 &        $-0.0205, -6.331$       &286.7, 5032.3       &        155.5        &         71.5        &        191      &        205      &        213 &        164      &        162 &        162 \\
        2 &        14.89 &        934.8 &        5.58 &        211.2 &        $-0.0529, -1.545$       &459.8, 2486.0       &        144.6        &         83.4        &        193      &        206      &        214 &        166      &        165 &        166 \\
        3 &        17.06 &        705.9 &        8.64 &        269.1 &        $-0.1262, -0.447$       &710.6, 1337.2       &        142.4        &         87.1        &        193      &        207      &        214 &        168      &        167 &        168 \\
\textbf{4}&\textbf{17.64}&\textbf{670.3}&\textbf{9.38}&\textbf{281.9}&\textbf{$-0.2291\pm i0$}      &\textbf{957.3}      &\textbf{142.2}       & \textbf{87.6}       &\textbf{194}     &\textbf{207}     &\textbf{214}&\textbf{169}     &\textbf{168}&\textbf{169}\\
        5 &        19.72 &        580.5 &       11.78 &        322.5 &       {$-0.1082\pm i0.1763$} &\textit{793.7}      &        141.7        &         88.7        &        197      &        208      &        213 &        170      &        169 &        170 \\
        6 &        22.83 &        500.8 &       14.95 &        373.8 &       {$-0.0289\pm i0.1832$} &\textit{654.7}      &        141.4        &         89.5        &        200      &        208      &        213 &        170      &        170 &        170 \\
        7 &        26.33 &        445.3 &       18.15 &        424.0 &       {$+0.0117\pm i0.1692$} &\textit{562.0}      &        141.2        &         90.0        &        202      &        208      &        212 &        171      &        171 &        171 \\
        8 &        30.20 &        404.4 &       21.37 &        473.4 &       {$+0.0344\pm i0.1536$} &\textit{496.0}      &        141.2        &         90.3        &        203      &        208      &        211 &        172      &        171 &        172 \\
\hline
\end{tabular}
\label{FitModParams}
\end{table*}

\begin{figure}[h]
    \centering
        \includegraphics[width=0.45\textwidth]{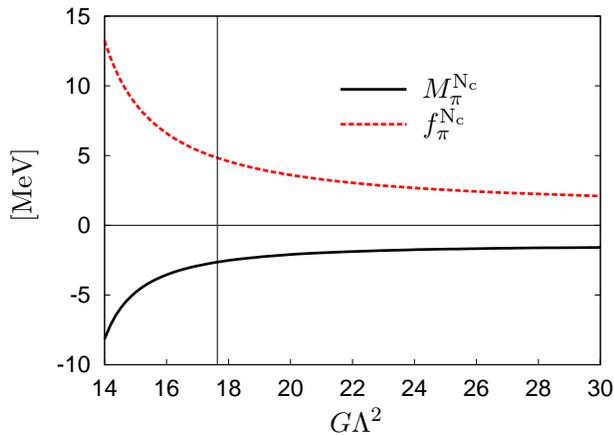}
  \caption{(Color online) $1/N_c$ corrections to the pion mass (black solid
line) and the weak pion decay constant (red dashed line).}
    \label{fpi}
\end{figure}

The mean-field contributions to the pion mass and weak decay constant are also
listed in Table~\ref{FitModParams} for the different parameter sets.
The corresponding $1/N_c$ corrections are shown in Fig.~\ref{fpi}.
For lower values of $G\Lambda^2$ these corrections amount to $-15$~MeV for the
pion mass and to $20$~MeV for $f_\pi$.
For parameter set 4 the corrections are only about $-2$~MeV and $5$~MeV, respectively.

The region of applicability of the model is closely related to the pole structure
of the mean-field quark propagator and depends on the parameters.
For large values of $\Lambda$, the propagator has two poles on the real
$p^2$ axis, whereas when $\Lambda$ is decreased these poles eventually merge and then
go over into a pair of complex conjugate poles.
(In both cases, there is an infinite number of additional complex conjugate poles,
which, however, only play a minor role.)
If the propagator has real poles, there is a real threshold in quark loop diagrams,
i.e., for external momenta $p>p_\mathrm{threshold}$ the loops have imaginary parts.
The complex conjugate poles, on the other hand, lead to a so-called pseudo-threshold.
In the absence of real poles the quark loops are then purely real but
have a cusp at the pseudothreshold.
Usually, the absence of poles on the real $p^2$ axis is considered as
a possible criterion for quark confinement \cite{Bhagwat:2002tx}.
However, due to the cusp in the real part of the quark loop diagrams
the applicability of the model above the pseudo-threshold is at least questionable.

\begin{figure}[h]
    \centering
        \includegraphics[width=0.45\textwidth]{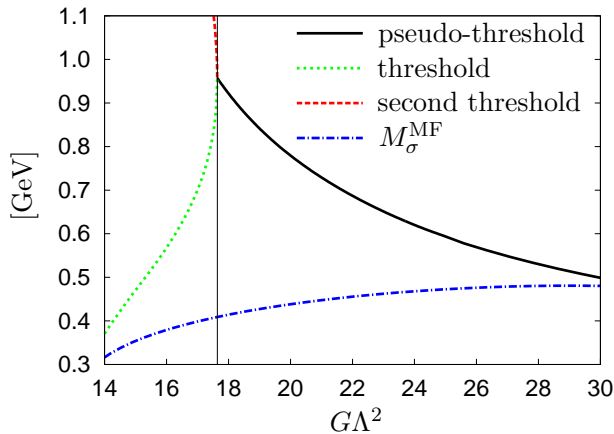}
  \caption{(Color online) The $\bar{q}q$-(pseudo)threshold due to the lowest
singularities of quark propagator.
Below a critical value of $G \Lambda^2$ (thin solid line) there are real
thresholds in quark loop diagrams (green dotted line for the first singularity
and red dashed line for the second one).
Above this critical value the lowest singularities of the quark propagator are
complex-conjugated values and there is no imaginary part of the quark loops.
However, quark loops contain a pseudo-threshold (thick solid line).
The blue dash-dotted line indicates the $\sigma$-meson mass at the mean field
level.}
    \label{Theshold}
\end{figure}

The positions of the two lowest quark poles and the corresponding
(pseudo-) thresholds of the polarization quark loop are listed in the table.
The (pseudo-) thresholds are also shown in Fig.~\ref{Theshold}.
There one can see that parameter set 4 corresponds to the maximal threshold value.
For this reason we will use parameter set 4 in the finite $T$ calculations
in Sec.~\ref{sec:T}.

\subsection{Sign of the $1/N_c$ corrections}
\label{sign}

Our results for the $1/N_c$-corrections to the quark condensate and to the pion
decay constant are quite surprising. Naively, one would expect that taking into account
meson loops should reduce the strength of spontaneous chiral symmetry breaking and,
thus, reduce the values of $|\ave{\bar qq}|$ and $f_\pi$. Indeed, this is what has been
found in Refs.~\cite{Oertel:1999fk,Oertel:2000jp} for the local NJL model.
However, as seen in Figs.~\ref{condensate} and \ref{fpi}, in the nonlocal model we
find exactly the opposite behavior for all sets of model parameters.
Obviously, this difference between the local and the nonlocal model calls for some
clarification.

As pointed out before, in the nonlocal model, after introducing the Gaussian form factor,
all diagrams at any order are automatically finite.
This is different from the local NJL model where, because of its non-renormalizability,
it is necessary to introduce independent cutoff parameters for pure quark loops and for
meson-quark loops, respectively.
In  \cite{Oertel:1999fk,Oertel:2000jp} a Pauli-Villars regularization has been
used for quark loops and a three-momentum cutoff $\Lambda_M$ for meson-quark
loops.
In order to study the transition from the nonlocal model to the local one let us
construct a nonlocal model with three parameters
\begin{enumerate}
\item parameter of nonlocality $\Lambda$
\item parameter of quark loop regularization $\Lambda_q$
\item parameter of meson loop regularization $\Lambda_M$
\end{enumerate}
The local model corresponds to the limit
\begin{eqnarray}
\Lambda\rightarrow\infty~,\quad
\Lambda_q^{}=\Lambda_q^{\rm fit}~,\quad
\Lambda_M^{}=\Lambda_M^{\rm fit}~,
\label{localLimit}
\end{eqnarray}
while the nonlocal model without regularization can be obtained by setting
\begin{eqnarray}
\Lambda=\Lambda^{\rm fit}~,\quad
\Lambda_q^{}\rightarrow\infty~,\quad
\Lambda_M^{}\rightarrow\infty~.
\label{nonlocalLimit}
\end{eqnarray}
For definiteness, let us compare the local model
\cite{Oertel:1999fk,Oertel:2000jp} with the nonlocal one from
\cite{Blaschke:2007np} with parameterizations fixed in the mean-field
approximation.
Note that for the given parameterizations, the MF quark condensates in the local
and the nonlocal models agree within less than $0.5~\%$.

\begin{figure}[!htb]
    \centering
    \includegraphics[width=0.45\textwidth]{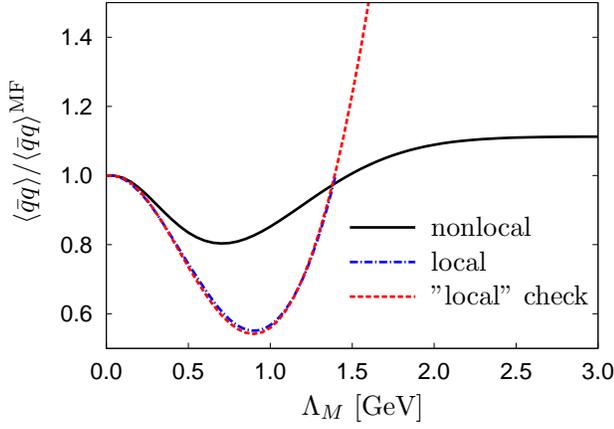}
\caption{(Color online) The ratio $\qq/\qq^{\mathrm{MF}}$ as a function of the
meson cutoff $\Lambda_M$. The full nonlocal result is shown by the black solid line.
The local result (blue dash-dotted line) is taken from Fig.~3a of
Ref.~\cite{Oertel:1999fk}. The "local" check (red dashed line)
is the local limit of the nonlocal calculations, see Eq.~(\ref{localLimit}).}
\label{Ratioqq}
\end{figure}

The next step is to consider the $1/N_c$ corrections and to investigate the
role of the mesonic three-momentum cut-off $\Lambda_M$.
For this purpose it is very instructive to study the ratio of the full quark
condensate to the MF contribution $\qq/\qq^{\mathrm{MF}}$.
In Fig.~\ref{Ratioqq} we compare the $\Lambda_M$ dependence of this ratio for
the local NJL model as given in Ref.~\cite{Oertel:1999fk} (dash-dotted line)
to that of the nonlocal model (bold solid line) and its local limit (dashed
line).\footnote{Note that in Ref.~\cite{Oertel:1999fk} the total quark condensate was
calculated including the perturbative part, whereas in the nonlocal model and its
local limit we have only considered the nonperturbative part. However, for the local
model the difference for nonstrange condensate is small.}
It is very interesting that in the region below $\sim 1.5$ GeV these models
predict a negative sign for the $1/N_c$ correction whereas for large mesonic
cut-off the sign is positive.
However, in the nonlocal model the absolute value of the correction saturates
for $\Lambda_M$ larger than $\sim 2.5$ GeV, which is well above actual
parameterizations for $\Lambda_q$ and $\Lambda_M$ in Ref.~\cite{Oertel:2000jp}.
In fact, in  Ref.~\cite{Oertel:1999fk} it was found that already for
$\Lambda_M \approx 1250$~MeV, the pion propagator gets unphysical poles, which was
interpreted as a breakdown of the expansion. In our present model no such
unphysical effects are observed.

\begin{figure}[!htb]
    \centering
    \includegraphics[width=0.45\textwidth]{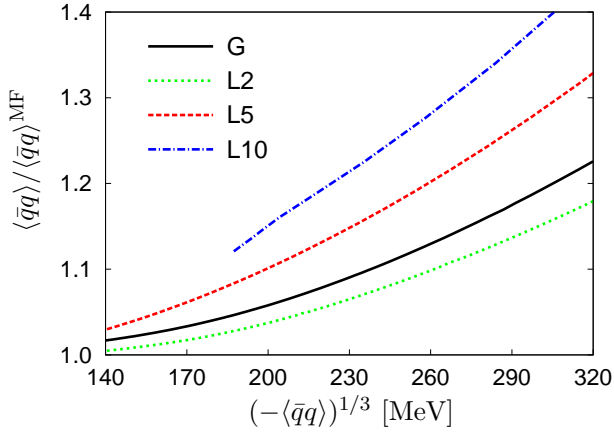}
\caption{(Color online) The ratio $\qq/\qq^{\mathrm{MF}}$ as a function of the
quark condensate for different form factors: Gaussian (G) and Lorentzian-type
(L2, L5, L10) with n=2, 5, 10. For details, see text.}
\label{RatiodiffFF}
\end{figure}

In order to study the dependence of the sign of the $1/N_c$ correction to the
quark condensate on the form factor we consider the Lorentzian-type form factor
$f(p_E^2)=1/(1+(p_E^2/\Lambda^2)^n)$ with n=2,5 or 10, see also
\cite{GomezDumm:2006vz,Grigorian:2006qe}.
We found that the sign of the $1/N_c$ correction is positive for all possible
parameterizations. This is shown in Fig.~\ref{RatiodiffFF}.

A related question concerns the $1/N_c$ corrections to the quark dressing
functions $A(p^2)$ and $B(p^2)$, defined by
\begin{equation}
    S^{-1}_p = A(p^2) \psl - B(p^2)\,,
\end{equation}
or, equivalently, to the wave-function renormalization function
$Z(p^2) = 1/A(p^2)$ and the mass function $M(p^2) = B(p^2)/A(p^2)$.
In mean-field approximation, we have $A^{MF}(p^2) \equiv 1$ and
$B^{MF}(p^2)= m_c + m_d f^2(p^2)$, see Eqs.~(\ref{Smf}) and (\ref{Sigmamf}).
Including $1/N_c$ corrections we find that the vector dressing function
$A(p^2)$ becomes enhanced, corresponding to a reduction of $Z(p^2)$.
However, the scalar dressing function $B(p^2)$ increases even stronger,
so that the mass function $M(p^2)$ is enhanced as well.

\begin{figure}[htb]
    \centering
    \includegraphics[width=0.45\textwidth]{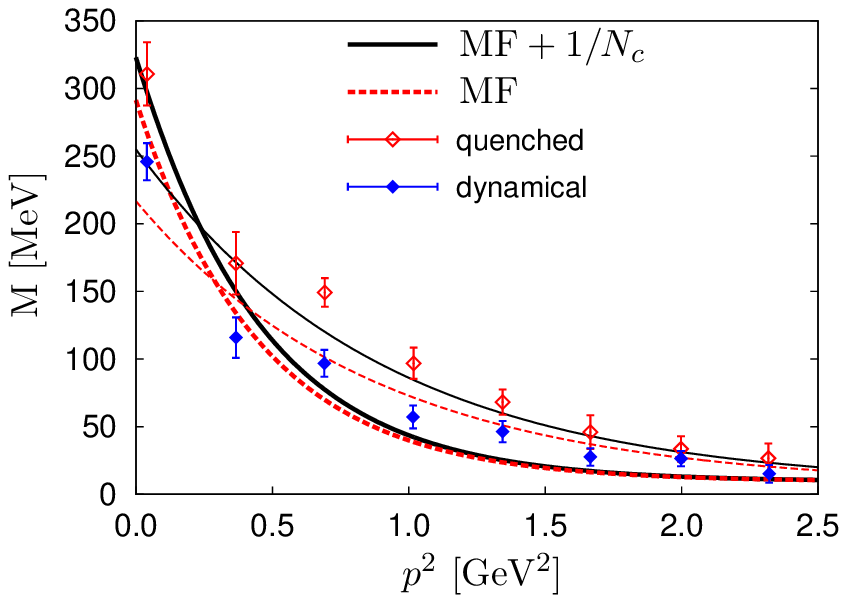}
    \includegraphics[width=0.45\textwidth]{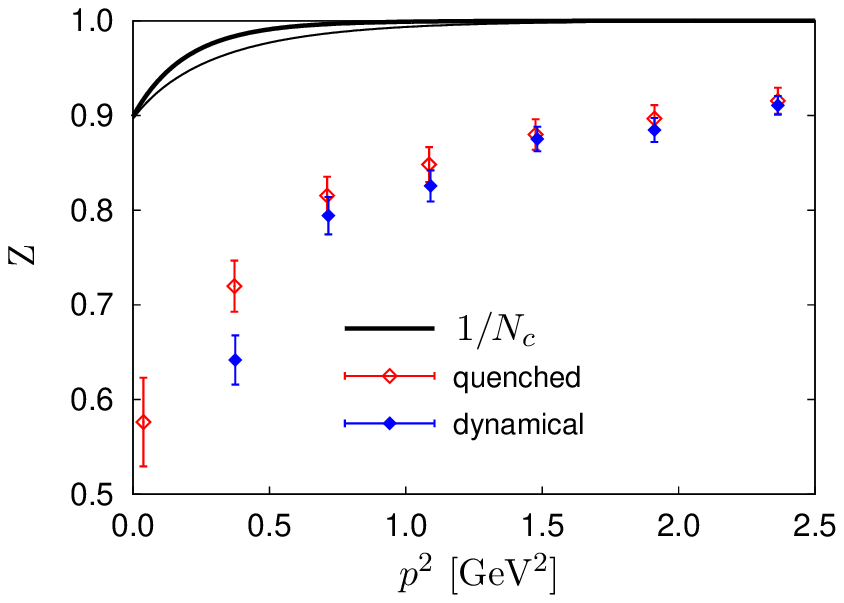}
\caption{(Color online) Behavior of the
quark mass function (upper panel) and wave-function renormalization function (lower panel)
for set No.~4 (bold lines) and set No.~2
(thin lines).
The results of the full $1/N_c$ corrected approach (black solid lines) and the
mean field contribution (red dashed lines) are compared with
quenched (open symbols) and unquenched (full symbols) lattice data
\cite{Kamleh:2007ud}, extrapolated to the chiral limit.}
\label{MZfunction}
\end{figure}

In Fig.~\ref{MZfunction} we present a comparison of the behavior of the
mass function (upper panel) and wave-function renormalization function
(lower panel) in the present approach for two parameter sets (No.~2 and No.~4)
with lattice data for this quantity \cite{Kamleh:2007ud} extrapolated to
the chiral limit case.
One can see that the gross behavior of the momentum dependence of the
quark mass function as measured in lattice QCD is reflected by the nonlocal
model. More in detail, the $1/N_c$ corrected result for the weaker coupling
No.~2 (thin black solid line)
matches the quark mass at zero-momentum, but stays above the
lattice data for dynamical quarks at finite momentum transfer.
The stronger coupling set No.~4 (bold black solid line), however,
overestimates the quark mass
function for low momenta $p^2<0.5$ GeV$^2$ by about 25\% and stays below the
lattice data by about the same amount in the intermediate momentum range
$0.5 \le p^2 [{\rm GeV}^2] \le 2$.

As one can see in the lower panel of Fig.~\ref{MZfunction}, the agreement
of our model results for the wave-function renormalization function $Z(p)$
with the lattice data is rather poor.
This is not unexpected because  Dyson-Schwinger QCD calculations find
a 30\% reduction of $Z(p)$ at low momenta already in quenched approximation,
i.e., without including meson effects~\cite{Fischer:2007ze},
whereas the present model predicts $Z\equiv 1$ at mean field.
A  nontrivial mean-field $Z(p)$ can be obtained in
the nonlocal PNJL model by adopting a suitable
separable form of the interaction with vector currents with derivative
coupling (rank-2 interaction), see
\cite{Noguera:2008,Contrera:2010,Horvatic:2010md,Hell:2011}
where also a comparison with lattice QCD data can be found.
In the present work which uses a separable ansatz in the
scalar-pseudoscalar channel (rank-one interaction), a nontrivial
wave-function renormalization arises only from the $1/N_c$ corrections.
Since these contributions generally do not exceed the 10\% level when
compared to the mean-field, we cannot expect a quantitative description of
$Z(p)$ at low momenta.

\begin{figure}[t]
    \centering
    \includegraphics[width=0.45\textwidth]{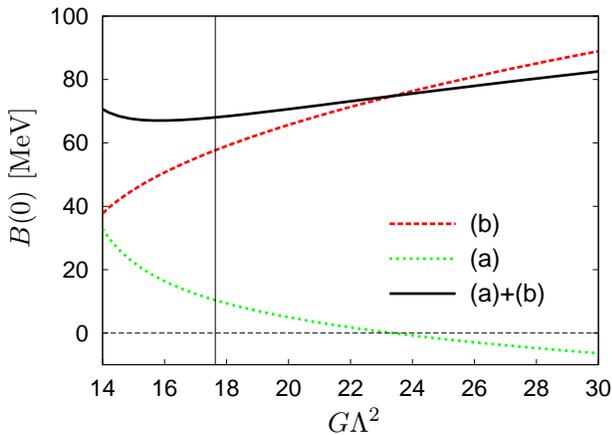}
\caption{(Color online) $1/N_c$-correction to the scalar quark dressing
         function $B(p^2)$ at $p^2 = 0$ as a function of the dimensionless
         coupling $G\Lambda^2$. Also shown are the separate contributions
         of the diagrams (a) and (b) of Fig.~\ref{Qucorrection}.}
\label{Bfunction}
\end{figure}

In Fig.~\ref{Bfunction} we show the $1/N_c$ correction to $B(0)$
as a function of the dimensionless coupling $G\Lambda^2$.
Besides the total correction (black solid line), which is almost independent
of the parameter choice, the contributions of the diagrams (a) and (b)
of Fig.~\ref{Qucorrection} are also indicated separately.
As one can see, the dominant contribution comes from diagram (b), which
is always positive.

Indeed, in Euclidean space the contribution of diagram (b) to the
$B$-function takes the form
\begin{alignat}{1}
    B^{N_c,(b)}(p_E^2)
     \,=&\, f^2(p_E^2)
      \int \frac{d^4 q_E}{(2\pi)^4}
      \frac{f^2(q_+^2)\, m(q_+^2)}{q_+^2 + m^2(q_+^2)}\times\,
\nonumber\\
&\times
      \left[
      \frac{3g_\pi^2(q_E^2)}{q_E^2+M_\pi^2} -
      \frac{g_\sigma^2(q_E^2)}{q_E^2+M_\sigma^2}
      \right]\,,
\label{BNcb}
\end{alignat}
where we have defined $q_+ = q_E+p_E$ and
\begin{equation}
    g_M^2(q_E^2) = \frac{q_E^2 + M_M^2}{G^{-1} - \Pi_M(q_E^2)}\,.
\end{equation}
The latter is a straightforward extension of Eqs.~(\ref{mesonprop}) and
(\ref{pionpole}) to (Euclidean) off-shell momenta.
In particular, $g_M^2$ is strictly positive.
Hence, the pion gives a positive contribution to $B^{N_c,(b)}(p_E^2)$, while
the sigma contribution is negative because of the extra minus sign.
Since the pions are more important, both, because of their lower mass
and because of the degeneracy factor of 3, the total contribution is positive,
as we have seen in Fig.~\ref{Bfunction}.

Neglecting the form factors and the momentum dependencies of the quark masses
and coupling constants, Eq.~(\ref{BNcb}) has a natural interpretation in terms
of a simple model where the quarks are coupled to pions and sigma mesons
by a Yukawa interaction,
\begin{equation}
      {\cal L}_\mathit{Yukawa} = -\bar q(x)\left[
               g_\sigma \sigma(x) + g_\pi i\gamma_5\vec\tau \cdot \vec\pi(x)
      \right] q(x)\,.
\end{equation}
From this point of view, the observed enhancement of the $B$-function
through pion loops should be the expected result.
This seems to be in conflict with the Dyson-Schwinger analysis of
Ref.~\cite{Fischer:2007ze}
where pion loops give a negative contribution to the $B$-function in QCD.
However, in that approach the pions have been introduced somewhat
differently, namely through corrections to the quark-gluon vertex.
It is interesting that this leads to the opposite sign.
Understanding this different behavior deserves further study.

At first sight, our results also seem to contradict the analysis of
Ref.~\cite{Muller:2010am} in a local NJL model,
where again it was found that the $B$-function decreases
when pion and sigma loops are included.
In contrast to our present model,
the $1/N_c$-corrections have been included selfconsistently, i.e.,
the calculations take into account the back-reactions of the meson loops
on the mean-field selfenergy. A closer inspection~\cite{Mueller} reveals
that this is the essential difference. Whereas, in complete agreement
with our expectations, the meson loops themselves give a positive contribution
to the $B$-function, their back-reaction strongly reduces the Hartree
contribution, so that the total effect is a reduction of $B(0)$.
In turn, this suggests that the enhancement of $B(0)$ we obtain in our model
could be due to the non-selfconsistent treatment of the $1/N_c$-corrections.
On the other hand, it is a well-known problem of the fully selfconsistent
scheme that the internal pions are much too heavy, so that in
Ref.~\cite{Muller:2010am} the (positive) contribution of the pion loops is
underestimated.
Hence, the true sign of the correction is the result of a delicate interplay
between several different processes and certainly needs further investigations.

\section{Finite temperature}
\label{sec:T}
In this section we extend the model to finite temperature and
then discuss the predictions for the pressure, the behavior of the quark
condensate and critical temperatures.

\subsection{Thermodynamic potential}

The model can easily be extended to finite temperature using a
$\Phi$-derivable ansatz (see, e.g., Ref.~\cite{Ripka:1997zb})
supplemented by the $1/N_c$ expansion.
The central quantity for the analysis is the thermodynamic potential per
volume
\begin{eqnarray}
\Omega =i\mathbf{Tr} \ln(\mathbf{S}^{-1}) + i \mathbf{Tr}(\Sigma \mathbf{S})
+ \Psi (\mathbf{S})+ U(\Phi,\bar{\Phi}) - \Omega_0~,
\nonumber\\
\label{Omega}
\end{eqnarray}
where $\mathbf{S}$ and $\Sigma = (S^c)^{-1} - \mathbf{S}^{-1}$ are the
full propagator and the quark selfenergy, respectively, and
$\mathbf{Tr}$ denotes the trace over all degrees of freedom,
internal ones and 4-momenta.
At nonzero temperature, we also take into account the Polyakov-loop dynamics
which no longer decouples from the quark sector.
To that end a constant temporal background gauge field
$\phi \equiv \ave{A_4} = \ave{iA_0}$ is minimally coupled to the quarks and
a Polyakov loop potential $U(\Phi,\bar{\Phi})$ is added in Eq.~(\ref{Omega}).
Here $\Phi = \frac{1}{N_c}\,\mathrm{Tr}_c\,e^{i\phi/T}$
denotes the Polyakov loop expectation value and $\bar{\Phi}$ its conjugate.
In order to avoid confusion with the potential of the $\Phi$-derivable scheme
(``$\Phi$ functional''), we denote the latter as $\Psi$.
In the exact case, $\Psi$ contains all two-particle irreducible diagrams.
Finally, we have introduced a subtractive renormalization constant $\Omega_0$,
which is chosen such that the vacuum ($T=0$) has vanishing pressure.

The thermodynamic equilibrium corresponds to the (global) minimum of
the thermodynamic potential with respect to the full quark propagator
and to the Polyakov loop, so that
the following necessary conditions (gap equations) must be fulfilled
\begin{eqnarray}
\frac{\partial \Omega}{\partial \mathbf{S}}=0,
\quad \frac{\partial \Omega}{\partial \Phi}=0~,
\quad \frac{\partial \Omega}{\partial \bar\Phi}=0~.
\label{gap}
\end{eqnarray}
We work in Polyakov gauge where the background gauge field
is diagonal in color space, i.e.,
$\phi = \phi_3\lambda_3 + \phi_8\lambda_8$.
Following ~\cite{Roessner:2006xn}, we require
$\Phi = \bar\Phi$ to be real with real $\phi_3,\phi_8$.
As a consequence $\phi_8=0$ and we are left with one variable $\phi_3$.
The second and the third equation thus reduce to
$\partial\Omega/\partial\phi_3 = 0$.
Moreover, the first equation implies that
$\Sigma = i{\delta \Psi}/{\delta\mathbf{S}}$.
Diagrammatically this means that the quark selfenergy can be obtained from
$\Psi$ by cutting one quark line at all possible places.

\begin{figure}[h]
    \centering
        \includegraphics[width=0.15\textwidth]{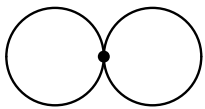}
  \caption{Glasses diagram for the $\Psi$ potential in the $\Phi$-derivable
scheme.}
    \label{glasses}
\end{figure}

Approximations can be introduced by truncating $\Psi$ at a certain order.
The mean-field results, corresponding to the leading order in $1/N_c$,
are obtained from the ``glasses'' diagram, displayed in Fig.~\ref{glasses}.
Solid lines represent dressed quark propagators.
In the nonlocal model the ``glasses'' potential takes the form
\begin{eqnarray}
\Psi_{\mathrm{glasses}} = -\sum\limits_{\mathrm{M}=\pi,\sigma}
\frac{G}{2}\left[-\mathbf{Tr}(\Gamma^{{\rm M}} i\mathbf{S})\right]^2,
\end{eqnarray}
and the thermodynamic potential reads
\begin{eqnarray}
\Omega^\mathrm{MF} = \frac{m_d^2}{2 G} +  U(\Phi,\bar{\Phi})
- \sum\limits_{i=0,\pm}4\int\limits_{k,n}\ln \left[k_{n,i}^2 + M^2(k_{n,i}^2) \right], \nonumber
\end{eqnarray}
where $k_{n,i}^2=(\omega_n^i)^2 +\mathbf{k}^2$ and the notation
$\int\limits_{k,n}\equiv T \sum\limits_n \int {d^3 k}/{(2\pi)^3}$ has been
introduced.
Note that due to the coupling to the Polyakov loop the fermionic Matsubara
frequencies $\omega_n=(2n+1)\pi T$ are partially shifted:
$\omega_n^\pm=\omega_n \pm \phi_3$ , $\omega_n^0=\omega_n$.
Apart from this shift, the modification of the quark propagator only
depends on the dynamical mass $m_d$. Therefore at mean field the gap
equations, Eq.~(\ref{gap}), take the form
\begin{eqnarray}
\frac{\partial \Omega^\mathrm{MF}}{\partial m_d}=0,
\qquad
\frac{\partial \Omega^\mathrm{MF}}{\partial\phi_3}=0.
\label{gapmf}
\end{eqnarray}

\begin{figure}[t]
    \centering
        \includegraphics[width=0.45\textwidth]{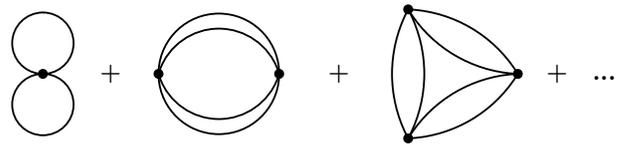}
  \caption{Ring sum in the $\Psi$-potential, see Eq.~(\ref{ringsum}).}
    \label{ring}
\end{figure}

The next-to-leading order contribution to the $\Psi$-potential is given
by the ``ring sum'',
\begin{eqnarray}
\Psi_{\mathrm{ring}} &=&
-\sum \limits_{\mathrm{M}=\pi,\sigma} \frac{d_\mathrm{M}}{2}
i\mathbf{Tr}\ln \left[ 1 - G \mathbf{\Pi}^{\mathrm{M}}\right],
\label{ringsum}
\end{eqnarray}
see Fig.~\ref{ring}.
Here $\mathbf{\Pi}^{\mathrm{M}}$ denotes the quark-antiquark polarization
functions constructed with
the full quark propagators and $d_M$ is the mesonic degeneracy factor.

At this level, the problem arises that in a fully selfconsistent treatment
the iteration of diagrams in the gap equation leads to contributions of
arbitrary orders in $1/N_c$. As a consequence, different approximation
schemes can be defined. In the present paper, we use a ``strict $1/N_c$
expansion'', where all contributions beyond the next-to-leading order
are discarded.
In the absence of the background gluon field this scheme is
straightforwardly implemented
by first solving the mean-field gap equation for $m_d$ (first equation
of Eq.~(\ref{gapmf})) and then evaluating the ring sum using the
mean-field propagators.
Thus, one gets for the thermodynamic potential
\begin{eqnarray}
    \Omega &=& \Omega^\mathrm{MF}+\Omega^\mathrm{N_c},
\label{Omegatot}
\end{eqnarray}
with the $1/N_c$ correction
\begin{eqnarray}
\label{Om_corr}
\Omega^\mathrm{N_c}&=&
\sum \limits_{M=\pi,\sigma} \frac{d_M}{2} \int\limits_{p,m}
\ln \left[ 1 - G \Pi_M(\vec{p},\nu_m)\right]
\end{eqnarray}
where the polarization functions $\Pi^{\mathrm{M}}$ are evaluated with
mean-field propagators and summation is over bosonic Matsubara frequencies.

Including the gluon background, we suggest to treat the Polyakov-loop
potential as effectively $N_c$ independent.\footnote{In principle
the $U(\Phi,\bar\Phi)$ is proportional to the number of gluons,
$N_c^2-1$~\cite{McLerran:2008ua,Pisarski:2000eq}.
Its leading contribution to the thermodynamic potential
is therefore of the order $O(N_c^2)$, while the
quarks only contribute at the order $O(N_c)$ and corrections are of the
order $O(N_c^0)$ for both, quarks and gluons.
However, since in practice the detailed form of $U$ is not based on
a $1/N_c$ expansion, but rather a phenomenological parameterization
fitted to quenched lattice data, we believe that it is more appropriate
to treat it as $N_c$ independent in the present context.}
A strict $1/N_c$ expansion of the thermodynamic potential then
corresponds to evaluate Eq.~(\ref{Omegatot}) for the simultaneous solutions
of the gap equations
\begin{eqnarray}
\frac{\partial \Omega^\mathrm{MF}}{\partial m_d}=0,
\qquad
\frac{\partial \Omega}{\partial\phi_3}=0.
\label{gapnc}
\end{eqnarray}
Note that $\phi_3$ is determined by minimizing the total thermodynamic
potential, whereas $m_d$ is obtained from the mean-field part only.
Nevertheless, since $\Omega^\mathrm{MF}$ also depends on $\phi_3$, the value
of $m_d$ is changed as well compared to the mean-field calculation,
due to the modified value of $\phi_3$.

We also note that the scheme outlined above slightly differs from the
scheme in our previous paper \cite{Blaschke:2007np} where both,
$m_d$ and $\phi_3$, have been fixed at mean-field level, before inserting
them into $\Omega^{N_c}$. In the numerical calculations, however,
this difference turned out to be small.

\subsection{Finite temperature results}

We now discuss our numerical results at finite temperature.
In the quark sector we take the parameters of set No.~4, see
Table~\ref{FitModParams} in Sec.~\ref{VacuumResults}.
For the Polyakov loop potential ${U}(\Phi,\bar{\Phi})$ we
adopt the logarithmic parametrization of Ref.~\cite{Roessner:2006xn}, which
has been fitted to the quenched lattice data of Ref.~\cite{Kaczmarek:2002mc}.
The only exception concerns the parameter $T_0$, which corresponds to the
transition temperature in pure gauge. While in Ref.~\cite{Roessner:2006xn}
the empirical value $T_0=270$~MeV was used, it has been argued in
Ref.~\cite{Schaefer:2007pw} that $T_0$ depends on the number of
active flavors and a value of $208$~MeV was suggested for $N_f=2$.
Therefore our calculations will be performed using both values of
$T_0$.

\begin{figure}[h]
    \centering
        \includegraphics[width=0.45\textwidth]{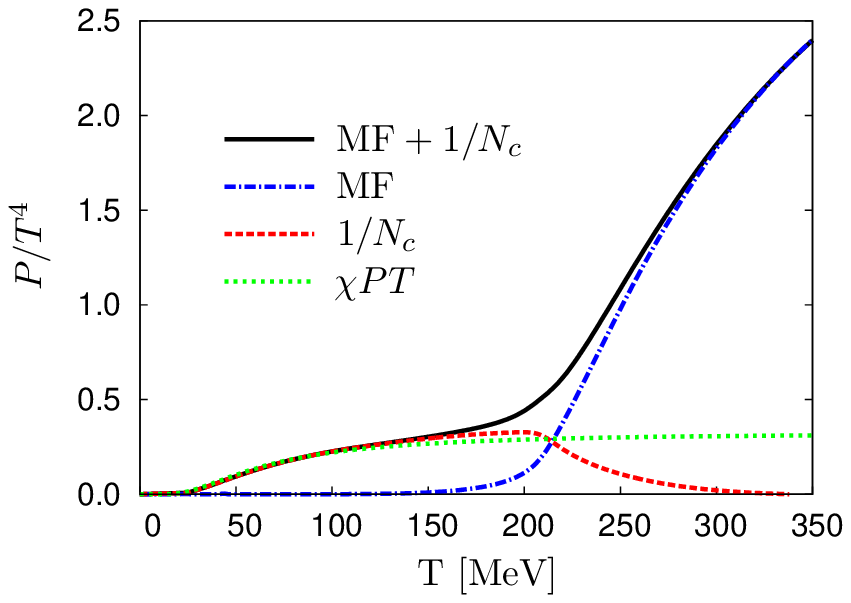}
        \includegraphics[width=0.45\textwidth]{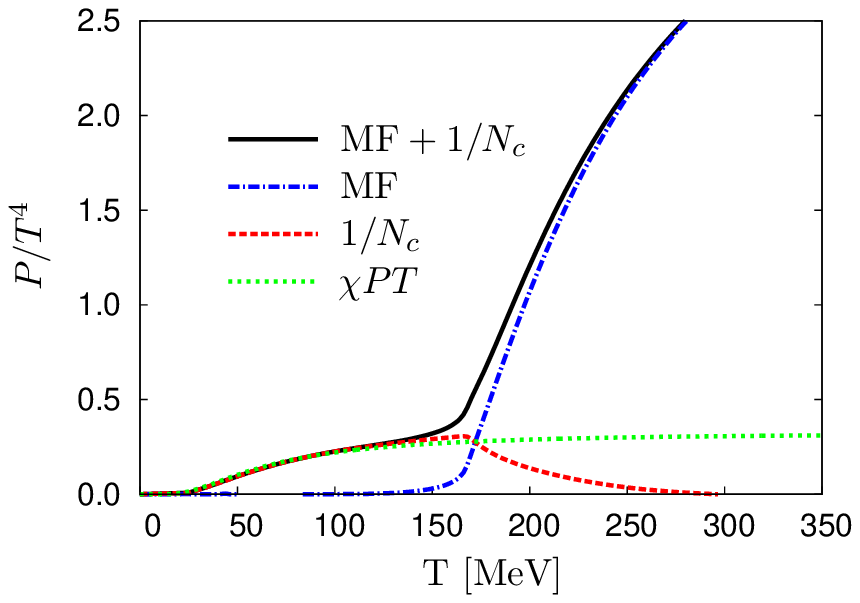}
  \caption{(Color online)  Temperature dependence of the scaled pressure
$P/T^4$ for $T_0 = 270$~MeV (upper panel) and for $T_0 = 208$~MeV
(lower panel):
mean field contribution (blue dash-dotted lines),
$1/N_c$ correction (red dashed lines),
mean field $+$ $1/N_c$ contributions (black solid lines).
The green dotted lines correspond to the
NLO chiral perturbation theory ($\chi PT$) result, Eq.~(\ref{PXPT})
\cite{Gasser:1986vb}.}
\label{Press}
\end{figure}

The model predictions for the pressure are displayed in Fig.~\ref{Press}.
The upper panel corresponds to $T_0 = 270$~MeV,
the lower one to $T_0 = 208$~MeV.
Together with the full result (solid lines) we show the partial contributions
to the pressure from the mean-field approximation (dash-dotted lines),
$P^\mathrm{MF}=-\Omega^\mathrm{MF}$, and from the $1/N_c$ corrections (dashed
lines), $P^\mathrm{N_c}=-\Omega^\mathrm{N_c}$.
In agreement with earlier results in PNJL-like models we find that
at low temperatures the mean-field contribution, corresponding to
thermally excited quarks, is strongly suppressed by the Polyakov loop.
In this regime the thermodynamics is governed by the ring sum,
which is dominated by
pionic degrees of freedom as the lightest particles in the mass spectrum.
Therefore, it is instructive to compare our result with the predictions
of chiral perturbation theory ($\chi PT$).  $\chi PT$ describes
the low-energy structure of different amplitudes in terms of an expansion in
powers of energies, momenta and current quark masses.
The finite temperature result for the pressure is given by
 \cite{Gasser:1986vb}
\begin{eqnarray}
\label{PXPT}
P_{\chi PT}
&=& \frac{N_f^2-1}{2}\left(g_0-\frac{1}{N_f} \frac{M_\pi^2}{2 F^2} g_1^2\right)
+O(p^8)~,
\nonumber\\
&&g_0 =-2T\int \frac{d^3p}{(2\pi)^3} \ln\left(1-e^{-E_\pi/T}\right)~,
\nonumber\\
&&g_1 =\int \frac{d^3p}{(2\pi)^3} \frac{1}{E_\pi\left(e^{E_\pi/T}-1\right)}~,
\end{eqnarray}
where $F$ is the weak pion decay constant in the chiral limit,
$E_\pi=\sqrt{\mathbf{p}^2+M_\pi^2}$ is the pion energy,
and $O(p^8)$ refers to the chiral counting scheme, where
$M_\pi$ and $T$ count as quantities of order $p$.
The term proportional to $g_0$ just represents the free relativistic pion gas
pressure, while the $g_1$ term is caused by interactions and
leads to a small reduction of the pressure.\footnote{In the chiral
limit $g_0$ and $g_1$ are given by
\begin{eqnarray}
g_0 =\frac{\pi^2 T^4}{45},\quad g_1 =\frac{T^2}{12}~~.
\end{eqnarray}}
The omitted terms of order $O(p^8)$ are also due to interactions.

Recalling that the pion decay constant is of the order $\sqrt{N_c}$,
we see that the $g_0$ and $g_1^2$ terms are of the order $N_c^0$ and
$1/N_c$, respectively.
Comparing this with our model, where the mean-field and ring-sum
contributions to the pressure are of the order $N_c$ and
$N_c^0$, respectively, we conclude that our model calculations
should only be consistent with the lowest-order $\chi PT$
result, i.e., with the $g_0$ term.
Nevertheless, since the $g_1^2$-term is small (it vanishes in the chiral
limit), we find excellent agreement even when this term is included,
see Fig.~\ref{Press}.
Moreover, the low-temperature behavior of our model predictions is almost
insensitive to the particular functional dependence of the form factor and
different parameterizations.
Of course, our results start to deviate from the $\chi PT$ predictions
when we approach the chiral phase transition.
Near the critical temperature the $\sigma$ meson gives an additional visible
contribution whereas already for $T>1.5~T_c$ the mesonic contributions are
negligible and the quark-gluon mean-field dominates the pressure.

Comparing the upper and lower panels of Fig.~\ref{Press} we observe that
the lowering of $T_0$ also leads to a lowering of the transition temperature.
However, apart from
this trivial effect, the results remain qualitatively unchanged.

\begin{figure}[b]
    \centering
        \includegraphics[width=0.45\textwidth]{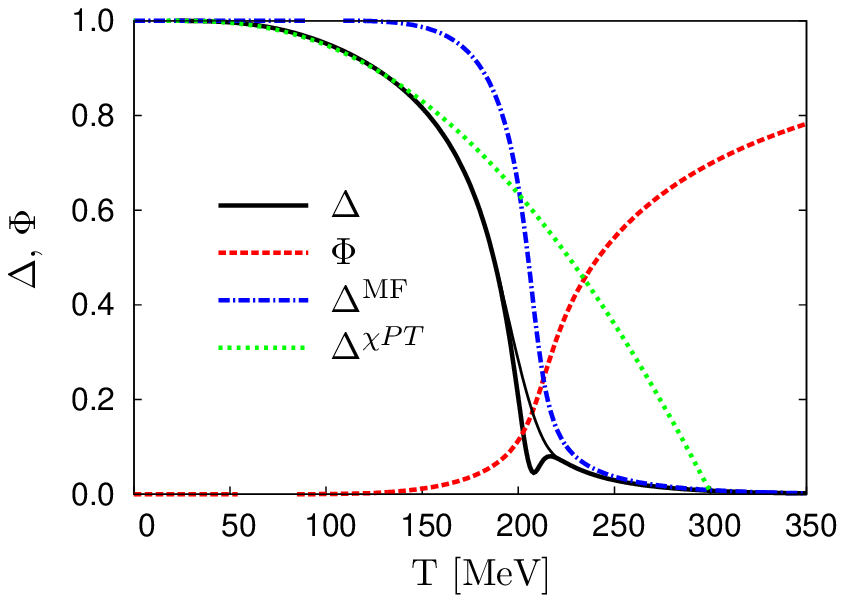}
        \includegraphics[width=0.45\textwidth]{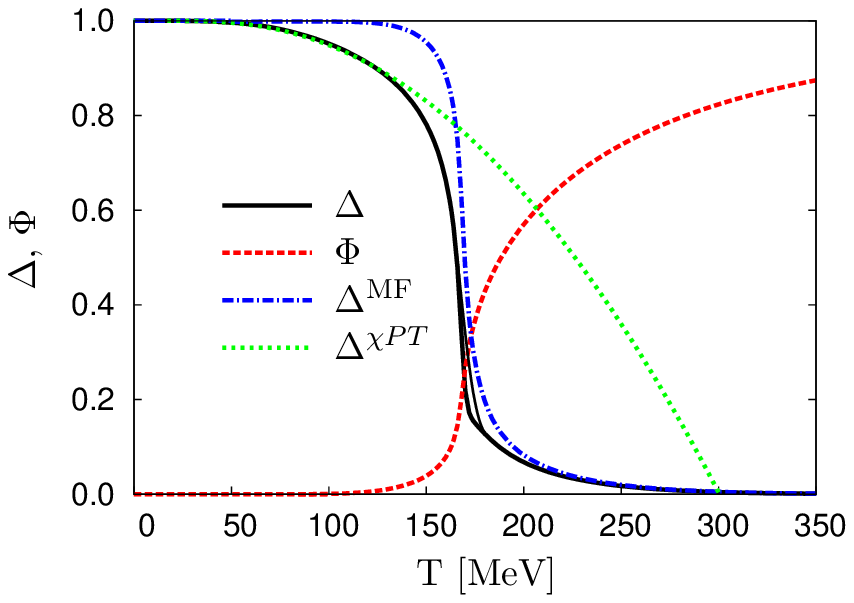}
  \caption{(Color online) Temperature dependence of the quark condensate
normalized to its vacuum value $\Delta=\qq_T/\qq$ (black solid lines) and the
Polyakov loop (red dashed lines) in the nonlocal PNJL model beyond mean field.
Furthermore shown are the mean-field result for the normalized quark
condensate $\Delta^{\rm MF}$ (blue dash-dotted lines),
the lowest-order chiral perturbation theory ($\chi PT$) result (green dotted
lines), and a na\"ive polynomial interpolation in the unstable regions of the
$1/N_c$ expansion (thin black solid lines).
The model calculations have been performed with $T_0 = 270$~MeV
(upper panel) and with $T_0 = 208$~MeV (lower panel).}
  \label{DynMF}
\end{figure}

In Fig.~\ref{DynMF} we show the temperature dependence of the quark condensate
$\langle\bar{q}q\rangle_T$ (black solid line)
and of the Polyakov loop expectation value (red dashed line)
in our model beyond mean field.
For comparison we also show the quark condensate in mean-field approximation
(blue dash-dotted line).
Again, the results for $T_0 = 270$~MeV and 208~MeV are displayed in the
upper and in the lower panel, respectively.
As already seen for the pressure, the $1/N_c$ corrections
mainly affect the behavior of the chiral condensate
below and around the critical region.
In particular at low temperatures the reduction of the chiral condensate
is almost entirely driven by the pion dynamics, which is missing in the
mean-field approximation.
In this context, it is again instructive to compare the model results
with the $\chi PT$ prediction. The latter is given by
\begin{eqnarray}
{\langle\bar{q}q\rangle_T} &=& {\langle\bar{q}q\rangle}
\left\{1-\frac{N_f^2-1}{N_f} \frac{g_1}{F^2} +O(p^4)\right\}~,
\end{eqnarray}
showing that the leading temperature effect is of the order $O(1/N_c)$.
Our model results are completely consistent with this: Whereas the
mean-field condensate ($O(1/N_c^0)$) stays practically constant at low
$T$, we find very good agreement with the $\chi PT$ predictions
(green dotted line) after including the $1/N_c$ corrections. We should note
that the chiral expansion scheme is very similar but not exactly equivalent
to the $1/N_c$ counting: In $\chi PT$ the leading-order correction to the
chiral condensate depends on the weak pion decay constant in the chiral
limit $F$, whereas in the $1/N_c$ expansion it depends on the
week pion decay constant for massive pions in mean-field approximation
$f_\pi^{\mathrm{MF}}$. Although formally of higher order in $p$ or $N_c$,
this could lead to quantitative differences, even at low temperatures.
However, for the chosen parameters the difference between $F$ and
$f_\pi^{\mathrm{MF}}$ happens to be very small.

In agreement with other works~\cite{Oertel:2000jp,Muller:2010am,Braun:2009si},
the additional reduction of the chiral condensate by the
$1/N_c$-correction terms generally also leads to a lowering of the
chiral phase-transition temperature as compared to the mean-field result.
This effect is most clearly seen in the upper panel of Fig.~\ref{DynMF},
corresponding to $T_0 = 270$~MeV.
Unfortunately, because of the perturbative nature of the strict
$1/N_c$-expansion scheme, our model cannot be applied to study the
phase transition itself. In the figure, this is obvious from the
existence of an unstable region (``wiggle'') in that region.
The origin of this wiggle can be attributed to the momentum independent
$1/N_c$ correction to the quark selfenergy (Fig.~\ref{Qucorrection}a).
In the vicinity of the phase transition, this diagram is dramatically
enhanced due to the restoration of chiral symmetry and the corresponding
lowering of the intermediate $\sigma$-meson mass.
(In fact, in the chiral limit this contribution would go to minus infinity
at $T_c$.)
As a rough estimate we define the unstable region as the regime where
the relative correction to the quark condensate is larger than $1/N_c$.
In the present example this corresponds to temperatures between
$183$ and $223$~MeV.
Since our results cannot be trusted in this area, we suggest to use a
simple polynomial interpolation between the stable regions at
lower and higher temperatures.
The resulting temperature dependence is displayed in Fig.~\ref{DynMF}
by the thin black solid line.

The situation is somewhat different for $T_0 = 208$~MeV (lower panel).
Although the $1/N_c$ corrections are again essential at low temperatures
(in agreement with the $\chi PT$ results), their effect on the chiral
critical temperature is almost negligible.
This is due to the fact that for this lower value of $T_0$ the mean-field
transition takes place earlier.
Related to this, the mean-field chiral condensate drops more
steeply\footnote{For $T_0 \lesssim 150$~MeV, the phase transition even
becomes first order. A similar observation was recently made in
Ref.~\cite{Horvatic:2010md}.}.
The ``unstable region'' where the relative correction to the quark
condensate is larger than $1/N_c$ is quite narrow
and the wiggle near the chiral phase transition, which we
found for $T_0 = 270$~MeV, is not present for $T_0 = 208$~MeV.

\begin{figure}[t]
    \centering
        \includegraphics[width=0.45\textwidth]{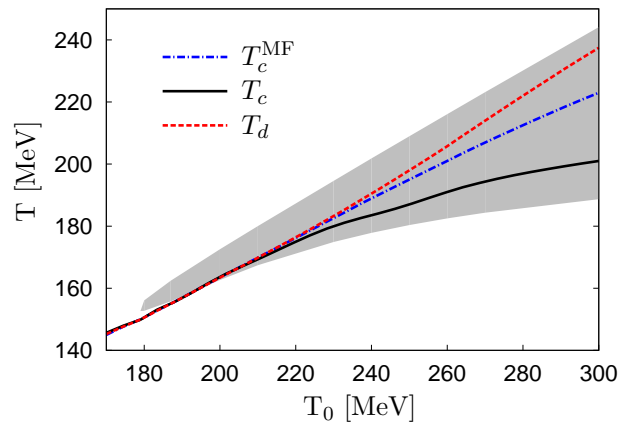}
  \caption{(Color online) Critical temperatures versus the parameter $T_0$
of the Polyakov-loop potential:
deconfinement (red dashed line), chiral restoration in
mean-field (blue dash-dotted line), chiral restoration with mesonic
fluctuations (black solid line).
The region with $|\langle\bar{q}q\rangle_T^{\mathrm{Nc}}/\langle\bar{q}q\rangle_T^{\mathrm{MF}}|>1/N_c$ is indicated by the shaded area.
}
\label{TcTd}
\end{figure}

A systematic overview about the $T_0$-dependence of the pseudocritical
temperatures for deconfinement (red dashed line) and chiral restoration
(solid black line) is given in Fig.~\ref{TcTd}.
As definitions of these temperatures we have chosen the maxima of the
temperature derivatives of the corresponding order parameter, i.e.,
Polyakov-loop expectation value and chiral condensate, respectively.
In the unstable regime (indicated by the shaded area) the latter was
obtained from the polynomial interpolation. The mean-field result for the
chiral-restoration temperature is also shown (blue dashed-dotted line).

In this figure, we can roughly distinguish three regimes:
At large values of $T_0$ already the mean-field $T_c$ is lower than
$T_d$, and $T_c$ is even lowered further by including $1/N_c$ corrections.
Then, in some intermediate regime, $T_d$ and $T_c$ agree at mean field but
there is still a visible reduction of the latter by the $1/N_c$ corrections.
Finally, for $T_0 \lesssim 220$~MeV all three lines practically
coincide.\footnote{According to Table~\ref{FitModParams} the $1/N_c$
corrections can even enhance the value of $T_c$ by 1 or 2~MeV at
$T_0 = 208$~MeV. However, since cross-over temperatures are not defined
in a unique way, and even rely on the polynomial interpolations in
the present case, this should not be over-interpreted. If we had defined
$T_c$ as the temperature where the condensate reaches half of its
vacuum value, the $1/N_c$ corrections would always lead to a reduction
of $T_c$.}

To understand this behavior we should recall that without coupling between
quark and Polyakov-loop sector,
$T_0$ corresponds to the (first-order) deconfinement-transition temperature,
while the chiral cross-over in the nonlocal model takes place at a very
low temperature, $T_c^{MF} = 116$~MeV.
From this point of view it is quite remarkable that even for $T_0=270$~MeV
the coupling reduces the difference between $T_c^{MF}$ and $T_d$ to
less than 10~MeV.
Nevertheless, the synchronization obviously works better when $T_0$ is
reduced and eventually becomes perfect for $T_0 \lesssim 220$~MeV.
Related to this, the transitions get sharper with decreasing $T_0$
and the relative effect of the $1/N_c$ corrections on $T_c$ is reduced
as well, as we have seen above.

At this point we should note that the Polyakov-loop dynamics and, hence,
the deconfinement transition are quite insensitive to the $1/N_c$
corrections. This is most likely an artifact of the model where the
Polyakov-loop is affected by the meson back-coupling only in a very
indirect way (see the discussion after Eq.~(\ref{gapnc})).
In principle, mesonic fluctuations should also directly influence the
Polyakov-loop potential. However, it is not clear how to
include this in PNJL-like models, where the potential is obtained from a
phenomenological fit of quenched lattice data where even the effects of
dynamical quarks are neglected (except for changes in $T_0$).
In this context it would certainly be interesting to compare the behavior
of the Polyakov loop with the so-called dressed Polyakov loop, related to
the dual quark condensate~\cite{Bilgici:2008qy,Fischer:2009wc}.
The latter can be calculated within PNJL-like~\cite{Kashiwa:2009ki}
(and even NJL models~\cite{Mukherjee:2010cp}) and it should be possible to
study $1/N_c$ corrections as well.
However, this is beyond the scope of the present paper.

\begin{figure}[t]
    \centering
        \includegraphics[width=0.45\textwidth]{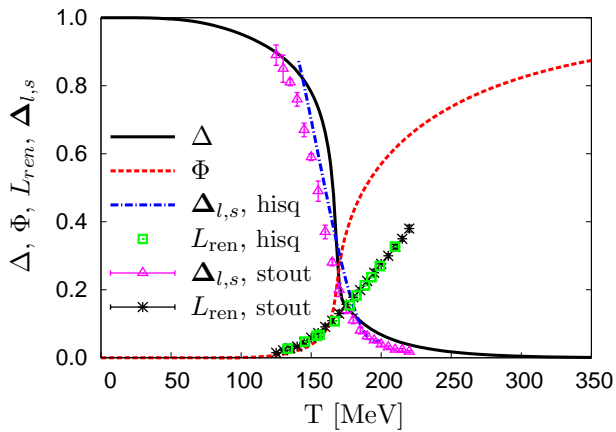}
  \caption{(Color online) Temperature dependence of the quark condensate
normalized to its vacuum value $\Delta=\qq_T/\qq$ (black solid line) and the
Polyakov loop (red dashed line) in the nonlocal PNJL model
with $T_0 = 208$~MeV beyond mean field
in comparison with lattice QCD data for the subtracted quark condensate
$\mathbf{\Delta}_{l,s}$ and for the renormalized Polyakov loop $L_\mathrm{ren}$.
Lattice points are taken from \cite{Bazavov:2010bx} (hisq action -- HotQCD
collaboration) and \cite{Borsanyi:2010bp} (stout action -- Wuppertal-Budapest
collaboration).
}
  \label{Dyn208Lat}
\end{figure}

In Fig.~\ref{Dyn208Lat} we present the comparison of our prediction for
the temperature dependence of quark condensate and Polyakov loop with recent
lattice data \cite{Bazavov:2010bx,Borsanyi:2010bp}.
The model calculations are performed with $T_0=208$~MeV, which leads to
the more realistic transition temperatures.
One should mention that the lattice calculations
\cite{Bazavov:2010bx,Borsanyi:2010bp} are performed for the case of two
light quarks and one strange, $N_f=2+1$.
For the investigation of the quark condensate they use a combination of
nonstrange($l$) and strange($s$) quark condensates, called subtracted quark
condensate,
\begin{eqnarray}
\mathbf{\Delta}_{l,s}=\frac{\langle\bar{q}q\rangle_T^{l}-\frac{m_c^l}{m_c^s}\langle\bar{q}q\rangle_T^{s}}
{\langle\bar{q}q\rangle^{l}-\frac{m_c^l}{m_c^s}\langle\bar{q}q\rangle^{s}}.
\label{subsqq}
\end{eqnarray}
In the present paper we consider a model with only two light active flavors.
Nevertheless, it makes sense to perform the comparison of our calculation for
the temperature dependence of the quark condensate normalized to its vacuum
value $\Delta=\qq_T/\qq$ with the lattice calculations of the subtracted quark
condensate.
The reason is that the nonpertubative part of the quark condensate in
Eq.~(\ref{qbarq}) is finite and the strange quark contribution in
Eq.~(\ref{subsqq}) is suppressed in the nonlocal model by two factors:
(i) the ratio of the nonpertubative part of the strange quark condensate to that
of the nonstrange one is $0.8$ and the ratio of the nonstrange current quark
mass to the strange one is about $1/20$.
On the other hand the main role for low temperature changes of the quark
condensate is played by pions - an effect which is correctly accounted for in
our model.

Indeed, we find that $\mathbf{\Delta}_{l,s}$ is rather well reproduced by our model
calculations for $\Delta$. For the Polyakov loop, on the other hand, we find
good agreement only at the onset, whereas around the deconfinement temperature,
$T \approx 170$~MeV, the model prediction rises much steeper than the lattice
data. This seems to indicate that there is missing physics, which is not
captured by just rescaling the temperature $T_0$ of the pure-glue potential.


\section{Summary}

In the present work, we have extended the nonlocal chiral quark model,
coupled to the Polyakov loop, beyond
the mean field approximation using a strict $1/N_c$ expansion scheme.
In vacuum, it is found that $1/N_c$ corrections lead to an increase of the
quark condensate, which is opposite to results obtained in the NJL model
with local interactions.
However, the result of the local model is strongly dependent on the mesonic
cut-off and for large values of the cut-off ($\Lambda_M> 1.5$  GeV) the sign
of the correction is positive in both models.

The parameters of the nonlocal model have been refitted in vacuum to reproduce
the physical values of the pion mass and the weak pion decay constant after
including the $1/N_c$ corrections.
In agreement with general expectations, we find that the mean field gives
the dominant contribution to the pion properties, while
the maximal size of the $1/N_c$ corrections to
the pion mass and weak pion decay constant amounts to 15 and 20 MeV,
respectively.

At finite temperature, the $1/N_c$ corrections lead to a reduction of
the chiral condensate when compared to the mean-field result.
Typically, this also leads to a reduction of the chiral phase transition
temperature.
However, for lower values of the parameter $T_0$ in the Polyakov-loop
potential, the mean-field transition becomes steeper and, thus, the effect
of the  $1/N_c$ corrections on $T_c$ becomes smaller.
Eventually, for $T_0 \lesssim 220$~MeV, $T_c$ remains practically unaffected
by the $1/N_c$ corrections.
For low temperatures, $T\leq 100$ MeV, our result for the quark condensate
practically coincides with that of $\chi PT$ whereas the high-temperature
region is well controlled by the mean-field approximation.

In the chiral limit our expansion scheme, which treats the $1/N_c$ corrections
perturbatively, breaks down in the vicinity of the chiral phase transition.
For the real case of nonzero current quark masses a similar situation takes
place for large values of the $T_0$ parameters of the Polyakov loop potential
(e.g., $T_0=270$ MeV in the pure gluodynamics case).
Alternatively, one could include the mesonic fluctuations non-perturbatively,
as done in Ref.~\cite{Muller:2010am} for the local NJL model.
In that analysis it was found as well that mesonic fluctuations lead to a
decrease of the critical temperature, but the unstable region around the
phase transition is absent.
On the other hand, the low-temperature behavior of $\chi PT$ is not
reproduced in this scheme.
It seems plausible that the inclusion of higher-order corrections could
make both expansion schemes converge to one another, i.e.,
in the strict $1/N_c$ expansion the unstable region will be reduced
(except close to a real phase transition) while in the self-consistent
scheme the low-temperature behavior will get closer to the $\chi PT$
predictions.

Concerning the pressure, we confirm our previous result \cite{Blaschke:2007np}
that the pionic contribution dominates in the low temperature region.
In this regime, the pressure is quite insensitive to the details of the
interaction and agrees almost exactly with that of an ideal pion gas.
At temperatures $T > T_c$, on the other hand, the mesonic contributions
die out at $T \sim 1.5~T_c$.

As a next step, we plan to study nonzero chemical potentials, which may require
a nontrivial extrapolation of the Polyakov-loop potential into this regime
\cite{Schaefer:2007pw}.
The treatment of the entire $T$--$\mu$ plane of the QCD phase diagram finally
requires the inclusion of baryonic degrees of freedom as well.

\begin{acknowledgments}
We thank C.~Fischer, D. M\"{u}ller, J.~Wambach, and R.~Williams for
critical remarks and illuminating discussions.
A.E.R. is grateful for the hospitality extended to him during visits at the
TU Darmstadt and at the University of Wroc{\l}aw.
We acknowledge support by the Heisenberg-Landau programme
(M.B., A.E.R., M.K.V.), by EMMI (A.E.R.),
by the Russian Foundation for Basic Research under
contracts No. 09-02-00749 (A.E.R.), No. 08-02-01003-a and No. 11-02-01538-a
(D.B.), a grant of the Russian President (A.E.R.),
and by the Polish Ministry of Science and Higher Education under
the CompStar-POL grant and grant No. NN 202 231837 (D.B.),
as well as by CompStar, a Research Networking Programme
of the European Science Foundation.
\end{acknowledgments}

\appendix

\begin{figure*}[!t]
\begin{tabular}{ccccccccccc}
\includegraphics[height=0.05\textheight]{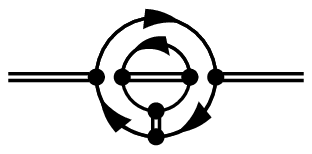} &\hspace{0.02\textwidth}&
\includegraphics[height=0.05\textheight]{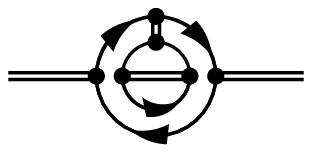}&\hspace{0.02\textwidth}&
\includegraphics[height=0.05\textheight]{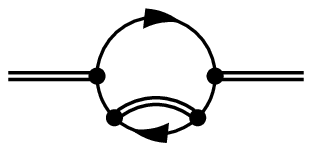} &\hspace{0.02\textwidth}&
\includegraphics[height=0.05\textheight]{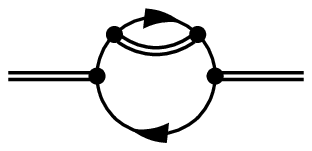}&\hspace{0.02\textwidth}&
\includegraphics[height=0.05\textheight]{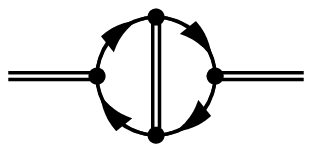} &\hspace{0.02\textwidth}&
\includegraphics[height=0.05\textheight]{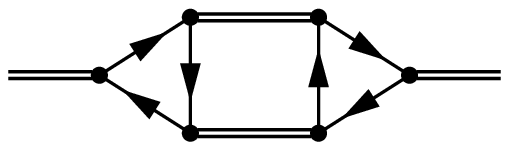}
\\ 1a && 1b && 2a && 2b && 2c && 3a
\end{tabular}
  \caption{\label{Meson1ncDiagram}Diagrams for the calculation of $1/N_c$ corrections to the meson
propagator.}
\end{figure*}

\begin{figure*}[!t]
\begin{tabular}{ccccccccc}
\includegraphics[height=0.05\textheight]{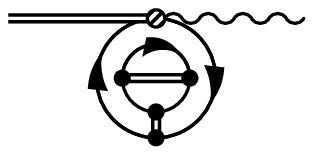} &\hspace{0.02\textwidth}&
\includegraphics[height=0.05\textheight]{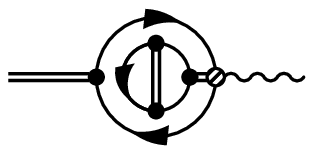} &\hspace{0.02\textwidth}&
\includegraphics[height=0.05\textheight]{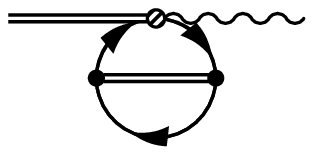} &\hspace{0.02\textwidth}&
\includegraphics[height=0.05\textheight]{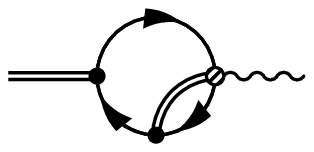}&\hspace{0.02\textwidth}&
\includegraphics[height=0.05\textheight]{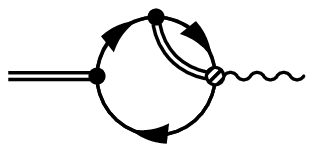}
\\
1c && 1d && 2d&&2e&&2f
\end{tabular}
\begin{tabular}{ccccccc}
\includegraphics[height=0.05\textheight]{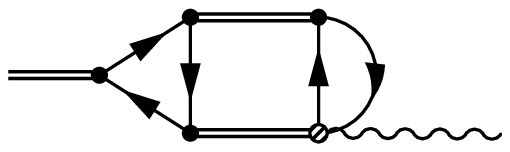} &\hspace{0.02\textwidth}&
\includegraphics[height=0.05\textheight]{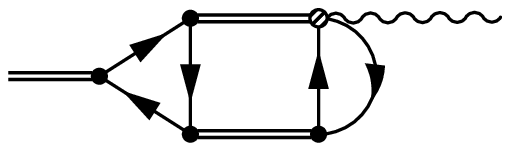}
\\
3b && 3c
\end{tabular}
  \caption{\label{fp1nc1Diagram}Additional nonlocal diagrams for the calculation of $1/N_c$
corrections to the weak pion decay.}
\end{figure*}

\section{$1/N_c$ corrections to the pion propagator and the
weak decay constant}
\label{appa}

The $1/N_c$ corrections to the meson propagator consist of the diagrams shown
in Fig.~\ref{Meson1ncDiagram}. For convenience we divide them into three types.
For the pion the corresponding expressions take the form
\begin{align}
&\Pi_{\pi,1a+2a}^{\mathrm{N_c}}(p^2)=
   i \intdk \times\nonumber\\
&\quad\quad\times \Tr
   \left[ \Gamma^{\pi}_{k_+,k_-}  S_{k_-}\Sigma^{\mathrm{N_c}}_{k_-} S_{k_-}\Gamma^{\pi}_{k_-,k_+}  S_{k_+} \right] \nonumber\\
&\Pi_{\pi,1b+2b}^{\mathrm{N_c}}(p^2)=
   i \intdk \times\nonumber\\
&\quad\quad\times\Tr
   \left[ \Gamma^{\pi}_{k_+,k_-}  S_{k_-}\Gamma^{\pi}_{k_-,k_+} S_{k_+} \Sigma^{\mathrm{N_c}}_{k_+} S_{k_+}\right] \nonumber
\end{align}
\begin{align}
&\Pi_{\pi,2c}^{\mathrm{N_c}}  (p^2)=
  \sum\limits_{\mathrm{M}=\sigma,\pi}
  \intdk \intdl \times \\
&\,\times\mathrm{Tr} \left[ \Gamma^{\pi}_{k_+,k_-} S_{k_-} \Gamma^\mathrm{M}_{k_-,l_-} S_{l_-} \Gamma^\pi_{l_-,l_+} S_{l_+} \Gamma^\mathrm{M}_{l_+,k_+} S_{k_+}\right] \mathrm{D}^{\mathrm{M}}_{k-l}\nonumber   \\
& \Pi_{\pi,3a}^{\mathrm{N_c}}  (p^2)=
  4 \times i \intdk
   \Gamma^{\pi\pi\sigma}_{k_+,k_-} \mathrm{D}^\pi_{k_-} \Gamma^{\pi\sigma\pi}_{k_-,k_+}  \mathrm{D}^\sigma_{k_+},\nonumber
\end{align}
where the factor $4$ in $\Pi_{\pi,3a}^{\mathrm{N_c}} (p^2)$ is the degeneracy
factor.

The calculation of the weak pion decay constant is more complicated.
Namely, part of the diagrams can be obtained
from the pion propagator times $-ig_\pi^\mathrm{MF}$ and substitution of the
outgoing pion vertex by the vertex with an external current.
Furthermore, there are additional nonlocal diagrams, shown in
Fig.~\ref{fp1nc1Diagram}.

The additional $1/N_c$ corrections of  type 1 are
\begin{eqnarray}
f_{\pi,1c}^{\mathrm{N_c}}  (p^2)&=&
 g_\pi^\mathrm{MF} \mathrm{C}  \intdk
   f^2(k) \Tr \left[  \Gamma^{5\pi ,L}_{k,k} S_k \Gamma^\mathrm{\sigma}_{k,k} S_k \right] \nonumber\\
f_{\pi,1d}^{\mathrm{N_c}}  (p^2)&=&
 g_\pi^\mathrm{MF} \mathrm{C} \intdk
  \Tr \left[  \Gamma^{\pi}_{k_+,k_-}  S_{k_-} \Gamma^{5\sigma ,L}_{k_-,k_+} S_{k_+}\right],    \nonumber
\end{eqnarray}

the $1/N_c$ corrections of  type 2 take the form
\begin{eqnarray}
&&f_{\pi,2d}^{\mathrm{N_c}}  (p^2)=
  g_\pi^\mathrm{MF} \intdk \intdl  \times\nonumber\\
   &&\quad\times\,\mathrm{Tr}
  \left[  \Gamma^{5\pi ,L}_{k,k} S_k \Gamma^\mathrm{M}_{k,l}  S_l \Gamma^\mathrm{M}_{l,k}  S_k \right] \mathrm{D}^\mathrm{M}_{k-l}\nonumber\\
&&f_{\pi,2e}^{\mathrm{N_c}}  (p^2)=
  g_\pi^\mathrm{MF} \intdk \intdl \times\nonumber\\
   &&\quad\times\,\mathrm{Tr}
  \left[ \Gamma^{\pi}_{k_+,k_-} S_{k_-} \Gamma^\mathrm{M}_{k_-,l_-} S_{l_-} \Gamma^{5M,L}_{l_-,k_+} S_{k_+}\right] \mathrm{D}^{\mathrm{M}}_{k-l}\nonumber\\
&&f_{\pi,2f}^{\mathrm{N_c}}  (p^2)=
  g_\pi^\mathrm{MF} \intdk \intdl \times\nonumber\\
   &&\quad\times\,\mathrm{Tr}
  \left[ \Gamma^{\pi}_{k_+,k_-} S_{k_-} \Gamma^\mathrm{5M,L}_{k_-,l_+} S_{l_+} \Gamma^\mathrm{M}_{l_+,k_+} S_{k_+}\right] \mathrm{D}^{\mathrm{M}}_{k-l}\nonumber
\end{eqnarray}
and should be summed over $\mathrm{M}=\pi,\sigma$.

The additional type-3 corrections are
\begin{eqnarray}
 f_{\pi,3b+3c}^{\mathrm{N_c}}  (p^2)&=&
   2 \times g_\pi^\mathrm{MF} i \intdk
   \Gamma^{\pi\pi\sigma}_{k_+,k_-} \mathrm{D}^\pi_{k_-} \Gamma^{5\sigma\pi,L}_{k_-,k_+}  \mathrm{D}^\sigma_{k_+},\nonumber
\end{eqnarray}
where the effective vertex $\Gamma^{5\sigma\pi,L}_{q_1,q_2}$ is

\begin{eqnarray}
\Gamma^{5\beta\gamma,L}_{q_1,q_2} &=&
 -  \intdl \Tr\,\left[ \Gamma^{5\gamma,L}_{l+q_1,l+q_2} S_{l+q_2}\Gamma^{\beta}_{l+q_2,l} S_l+\right.\nonumber\\
&&\quad\left. +\Gamma^{5\beta,L}_{l+q_1,l+q_2}  S_l \Gamma^{\gamma}_{l,l+q_1} S_{l+q_1}\right].
\label{triangleEffAxialVertex}
\end{eqnarray}


\begin{thebibliography}{99}
\bibitem{Nambu:1961tp}
Y.~Nambu and G.~Jona-Lasinio,
Phys.\ Rev.\ {\bf 122}, 345 (1961);
Phys.\ Rev.\ {\bf 124}, 246 (1961).

\bibitem{Volkov:1984kq}
M.~K.~Volkov,
Annals Phys.\ {\bf 157}, 282 (1984);
Sov. J. Part. Nucl. {\bf 17}, 186 (1986).

\bibitem{Klimt:1989pm}
S.~Klimt, M.~Lutz, U.~Vogl and W.~Weise,
Nucl.\ Phys.\ A {\bf 516}, 429 (1990);
Nucl.\ Phys.\ A {\bf 516}, 469 (1990).

\bibitem{Klevansky:1992qe}
S.~P.~Klevansky,
Rev.\ Mod.\ Phys.\ {\bf 64}, 649 (1992).

\bibitem{Hatsuda:1994pi}
T.~Hatsuda and T.~Kunihiro,
Phys.\ Rept.\ {\bf 247}, 221 (1994).

\bibitem{Meisinger:1995ih}
P.~N.~Meisinger and M.~C.~Ogilvie,
Phys.\ Lett.\ B {\bf 379}, 163 (1996).

\bibitem{Fukushima:2003fw}
K.~Fukushima,
Phys.\ Lett.\ B {\bf 591}, 277 (2004).

\bibitem{Megias:2004hj}
E.~Megias, E.~Ruiz Arriola and L.~L.~Salcedo,
Phys.\ Rev.\ D {\bf 74}, 065005 (2006).

\bibitem{Ratti:2005jh}
C.~Ratti, M.~A.~Thaler, and W.~Weise,
Phys.\ Rev.\ D {\bf 73}, 014019 (2006).

\bibitem{Roessner:2006xn}
S.~R\"o{\ss}ner, C.~Ratti, and W.~Weise,
Phys.\ Rev.\ D {\bf 75}, 034007 (2007).

\bibitem{Sasaki:2006ww}
C.~Sasaki, B.~Friman and K.~Redlich,
Phys.\ Rev.\  D {\bf 75}, 074013 (2007).

\bibitem{Hansen:2006ee}
H.~Hansen, W.~M.~Alberico, A.~Beraudo, A.~Molinari, M.~Nardi, and C.~Ratti,
Phys.\ Rev.\ D {\bf 75}, 065004 (2007).

\bibitem{Blaschke:2007np}
D.~Blaschke, M.~Buballa, A.~E.~Radzhabov and M.~K.~Volkov,
Phys.\ Atom.\ Nucl.\ {\bf 71}, 1981 (2008).

\bibitem{Contrera:2007wu}
G.~A.~Contrera, D.~Gomez Dumm and N.~N.~Scoccola,
Phys.\ Lett.\ B {\bf 661}, 113 (2008).

\bibitem{Hell:2009by}
T.~Hell, S.~R\"o{\ss}ner, M.~Cristoforetti and W.~Weise,
Phys.\ Rev.\ D {\bf 81}, 074034 (2010).

\bibitem{Horvatic:2010md}
D.~Horvatic, D.~Blaschke, D.~Klabucar and O.~Kaczmarek,
arXiv:1012.2113 [hep-ph].

\bibitem{Kondo:2010ts}
K.~I.~Kondo,
Phys.\ Rev.\ D {\bf 82}, 065024 (2010).

\bibitem{Kamleh:2007ud}
W.~Kamleh, P.~O.~Bowman, D.~B.~Leinweber, A.~G.~Williams, J.~Zhang,
Phys.\ Rev.\  {\bf D76}, 094501 (2007).

\bibitem{Bazavov:2010bx}
A.~Bazavov and P.~Petreczky [HotQCD Collaboration],
arXiv:1009.4914 [hep-lat].

\bibitem{Borsanyi:2010bp}
S.~Borsanyi, Z.~Fodor, C.~Hoelbling, S.~D.~Katz, S.~Krieg, C.~Ratti and K.~K.~Szabo
[Wuppertal-Budapest Collaboration],
JHEP {\bf 1009}, 073 (2010).

\bibitem{Braun:2009gm}
J.~Braun, L.~M.~Haas, F.~Marhauser and J.~M.~Pawlowski,
Phys.\ Rev.\ Lett.\ {\bf 106}, 022002 (2011).

\bibitem{Amsler:2008zzb}
C.~Amsler {\it et al.} [Particle Data Group],
Phys.\ Lett.\ B {\bf 667}, 1 (2008).

\bibitem{Quack:1993ie}
E.~Quack and S.~P.~Klevansky,
Phys.\ Rev.\ C {\bf 49}, 3283 (1994).

\bibitem{Ebert:1994sm}
D.~Ebert, M.~Nagy, and M.~K.~Volkov,
Phys.\ Atom.\ Nucl.\ {\bf 59}, 140 (1996).

\bibitem{Nikolov:1996jj}
E.~N.~Nikolov, W.~Broniowski, C.~V.~Christov, G.~Ripka, and K.~Goeke,
Nucl.\ Phys.\ A {\bf 608}, 411 (1996).

\bibitem{Dmitrasinovic:1995cb}
V.~Dmitrasinovic, H.~J.~Schulze, R.~Tegen and R.~H.~Lemmer,
Annals Phys.\ {\bf 238}, 332 (1995).

\bibitem{Blaschke:1995gr}
D.~Blaschke, Yu.~L.~Kalinovsky, G.~R\"opke, S.~M.~Schmidt and M.~K.~Volkov,
Phys.\ Rev.\ C {\bf 53}, 2394 (1996).

\bibitem{Oertel:1999fk}
M.~Oertel, M.~Buballa and J.~Wambach,
Phys.\ Lett.\ B {\bf 477}, 77 (2000).

\bibitem{Oertel:2000jp}
M.~Oertel, M.~Buballa and J.~Wambach,
Phys.\ Atom.\ Nucl.\ {\bf 64}, 698 (2001).

\bibitem{Plant:2000ty}
R.~S.~Plant and M.~C.~Birse,
Nucl.\ Phys.\ A {\bf 703}, 717 (2002).

\bibitem{Jafarov:2003pe}
R.~G.~Jafarov and V.~E.~Rochev,
Central Eur.\ J.\ Phys.\ {\bf 2}, 367 (2004).

\bibitem{Goeke:2007bj}
K.~Goeke, M.~M.~Musakhanov and M.~Siddikov,
Phys.\ Rev.\ D {\bf 76}, 076007 (2007).

\bibitem{Muller:2010am}
D.~M\"uller, M.~Buballa and J.~Wambach,
Phys.\ Rev.\ D {\bf 81}, 094022 (2010).

\bibitem{'tHooft:1973jz}
G.~'t Hooft,
Nucl.\ Phys.\ B {\bf 72}, 461 (1974).

\bibitem{GomezDumm:2006vz}
D.~Gomez Dumm, A.~G.~Grunfeld, N.~N.~Scoccola,
Phys.\ Rev.\  {\bf D74}, 054026 (2006).

\bibitem{Terning:1991yt}
J.~Terning,
Phys.\ Rev.\ D {\bf 44}, 887 (1991).

\bibitem{Bowler:1994ir}
R.~D.~Bowler and M.~C.~Birse,
Nucl.\ Phys.\ A {\bf 582}, 655 (1995).

\bibitem{Plant:1997jr}
R.~S.~Plant and M.~C.~Birse,
Nucl.\ Phys.\ A {\bf 628}, 607 (1998).

\bibitem{Dorokhov:2003kf}
A.~E.~Dorokhov and W.~Broniowski,
Eur.\ Phys.\ J.\ C {\bf 32}, 79 (2003).

\bibitem{Scarpettini:2003fj}
A.~Scarpettini, D.~Gomez Dumm and N.~N.~Scoccola,
Phys.\ Rev.\ D {\bf 69}, 114018 (2004).

\bibitem{Bhagwat:2002tx}
M.~Bhagwat, M.~A.~Pichowsky and P.~C.~Tandy,
Phys.\ Rev.\ D {\bf 67}, 054019 (2003).

\bibitem{Grigorian:2006qe}
H.~Grigorian,
Phys.\ Part.\ Nucl.\ Lett.\ {\bf 4}, 223 (2007).

\bibitem{Fischer:2007ze}
C.~S.~Fischer, D.~Nickel, J.~Wambach,
Phys.\ Rev.\  {\bf D76}, 094009 (2007).

\bibitem{Noguera:2008}
S.~Noguera and N.~N.~Scoccola,
Phys.\ Rev.\ D {\bf 78}, 114002 (2008).

\bibitem{Contrera:2010}
G.~A.~Contrera, M.~Orsaria, and N.~N.~Scoccola,
Phys.\ Rev.\ D {\bf 82}, 054026 (2010).

\bibitem{Hell:2011}
T.~Hell, K.~Kashiwa, and W.~Weise,
arxiv:1104.0572 [hep-ph].

\bibitem{Mueller}
D.~M\"uller, private communication.

\bibitem{Ripka:1997zb}
G.~Ripka,
``Quarks bound by chiral fields:
The quark-structure of the vacuum and of light mesons and baryons,''
{\it Oxford, UK: Clarendon Pr. (1997).}

\bibitem{McLerran:2008ua}
L.~McLerran, K.~Redlich and C.~Sasaki,
Nucl.\ Phys.\ A {\bf 824}, 86 (2009).

\bibitem{Pisarski:2000eq}
R.~D.~Pisarski,
Phys.\ Rev.\ D {\bf 62}, 111501 (2000).

\bibitem{Kaczmarek:2002mc}
O.~Kaczmarek, F.~Karsch, P.~Petreczky and F.~Zantow,
Phys.\ Lett.\ B {\bf 543}, 41 (2002).

\bibitem{Schaefer:2007pw}
B.-J.~Schaefer, J.~M.~Pawlowski and J.~Wambach,
Phys.\ Rev.\ D {\bf 76}, 074023 (2007).

\bibitem{Gasser:1986vb}
J.~Gasser and H.~Leutwyler,
Phys.\ Lett.\ B {\bf 184}, 83 (1987).

\bibitem{Braun:2009si}
J.~Braun,
Phys.\ Rev.\ D {\bf 81}, 016008 (2010).

\bibitem{Bilgici:2008qy}
E.~Bilgici, F.~Bruckmann, C.~Gattringer and C.~Hagen,
Phys.\ Rev.\  D {\bf 77}, 094007 (2008).

\bibitem{Fischer:2009wc}
C.~S.~Fischer,
Phys.\ Rev.\ Lett.\  {\bf 103}, 052003 (2009).

\bibitem{Kashiwa:2009ki}
K.~Kashiwa, H.~Kouno and M.~Yahiro,
Phys.\ Rev.\  D {\bf 80}, 117901 (2009).

\bibitem{Mukherjee:2010cp}
T.~K.~Mukherjee, H.~Chen and M.~Huang,
Phys.\ Rev.\  D {\bf 82}, 034015 (2010).

\end{thebibliography}
\end{document}